\documentclass[showpacs, preprintnumbers, nofootinbib, aps, prd, superscriptaddress,10pt, showkeys, notitlepage, twocolumn]{revtex4-1}

\usepackage{graphicx,amssymb,amsmath,amsthm,amsfonts,epsfig, setspace}

\usepackage[linktocpage]{hyperref}
\usepackage[usenames,dvipsnames]{color}
\usepackage{epstopdf}
\usepackage{aas_macros}
\usepackage{pifont}
\definecolor{darkred}{rgb}{0.5,0,0}
\definecolor{darkgreen}{rgb}{0,0.5,0}
\definecolor{darkblue}{rgb}{0,0,0.5}
\definecolor{prussian}{rgb}{0.0, 0.19, 0.33}
\definecolor{richelectricblue}{rgb}{0.03, 0.57, 0.82}
\definecolor{teal}{rgb}{0.0, 0.5, 0.5}
\definecolor{mediumseagreen}{rgb}{0.24, 0.7, 0.44}
\definecolor{lust}{rgb}{0.9, 0.13, 0.13}
\definecolor{ballblue}{rgb}{0.13, 0.67, 0.8}
\definecolor{darkcyan}{rgb}{0.0, 0.55, 0.55}
\definecolor{mountainmeadow}{rgb}{0.19, 0.73, 0.56}
\definecolor{palecarmine}{rgb}{0.69, 0.25, 0.21}
\definecolor{richcarmine}{rgb}{0.84, 0.0, 0.25}
\definecolor{tangelo}{rgb}{0.98, 0.3, 0.0}
\definecolor{venetian}{rgb}{0.784,0.031,0.082}
\definecolor{bdfrance}{rgb}{0.192,0.549,0.906}

\hypersetup{colorlinks=true, citecolor=venetian,
linkcolor=bdfrance, urlcolor=lust}
\usepackage{amsmath,amssymb}
\usepackage{tensor}
\usepackage{mathtools}
\usepackage{amsbsy}
\usepackage{bm}
\usepackage{float}

\newcommand{\be}{\begin{equation}}
\newcommand{\ee}{\end{equation}}
\newcommand{\bear}{\begin{eqnarray}}
\newcommand{\eear}{\end{eqnarray}}

\newcommand{\p}{\prime}
\newcommand{\pp}{\prime\prime}
\newcommand{\nn}{\nonumber}

\newcommand{\cD}{{\cal D}}
\newcommand{\cS}{{\cal S}}

\newcommand{\cM}{{\cal M}}

\newcommand{\rS}{{\rm s}}
\newcommand{\rK}{{\rm K}}

\newcommand{\rph}{r_{\rm ph}}

\newcommand{\bph}{b_{\rm ph}}

\begin{document}

\title{Is a black hole shadow a reliable test of the no-hair theorem?}

\author{Kostas Glampedakis}
\email{kostas@um.es}
\affiliation{Departamento de F\'isica, Universidad de Murcia, Murcia, E-30100, Spain}
\affiliation{Theoretical Astrophysics, University of T\"ubingen, Auf der Morgenstelle 10, T\"ubingen, D-72076, Germany}

\author{George Pappas}
\email{gpappas@auth.gr}
\affiliation{Department of Physics, Aristotle University of Thessaloniki, Thessaloniki 54124, Greece}

\begin{abstract}
Capturing the image of the shadow cast by the event horizon of an illuminated black hole is, at the most basic level, an experiment of extreme light deflection
in a strongly curved spacetime. As such, the properties of an imaged shadow can be used to probe the general relativistic Kerr nature of astrophysical black 
holes. As an example of this prospect, it is commonly asserted that a shadow can test the validity of the theory's famous `no hair theorem' for the black hole's 
mass and spin multipole moments. In this paper, we assess this statement by calculating the shadow's equatorial radius 
in spacetimes with an arbitrary multipolar structure and within a slow rotation approximation. We find that when moments higher than the quadrupole are taken 
into account, the shadow acquires a high degree of degeneracy as a function of the deviation from the Kerr multipole moments.  The results of our analysis suggest that dark 
objects with strongly non-Kerr multipolar structure could nevertheless produce a Kerr-like shadow with its characteristic quasi-circular shape.

\end{abstract} 
  
\maketitle


\section{Introduction}
\label{sec:intro}

The last ten years or so have seen a revolution in the ways we probe strong-field relativistic gravity. The main breakthrough came in 2015  
with the first observation of gravitational waves (GWs) from merging black holes by the LIGO-Virgo Collaboration~\cite{Abbott:2016blz} . 
Since then these detectors have chalked up many more merging compact binary systems, allowing for new precision tests of General Relativity (GR)
and new astrophysical information on compact objects (for a review, see~\cite{isi2021}). 
The second most important milestone, and the one most relevant to this paper, was reached by the spectacular high-resolution 
image of the supermassive black hole in the M87 galactic center (usually dubbed M87*), obtained by the Event Horizon Telescope (EHT) 
Collaboration~\cite{EHT2019ApJ.V}. During the preparation of this paper a second sister image was released, picturing the  supermassive black hole SgrA*
in our galactic center~\cite{EHT22}. These images serve as direct evidence for the existence of black holes and can also be used
as probes of GR  (for a review, see, e.g., Ref.~\cite{Broderick2014ApJ}). Motivated by this exciting possibility,  a significant amount of work over the last decade or so, 
has focused on the calculation of shadows of black holes beyond GR 
(e.g., Refs.~\cite{Johannsen2010ApJ, Johannsen2013ApJ, Cunha2015PhRvL, Cunha2016IJMPD, Cunha2016PhRvD, 
Cunha2017PhysRevD, Cunha2017PhysRevLett, Cunha2018GReGr, Medeiros:2019cde, wielgus21, junior21, lara21})  
as well as on improving our understanding of the image produced by general relativistic Kerr black holes (e.g., Refs.~\cite{Gralla_etal2019, Gralla2020, 
Gralla_Lupsasca2020, bronzwaer21, paugnat22}).

The centerpiece in these images is the shadow cast by the black hole as silhouetted against its luminous accretion flow. The shape and overall scale of the shadow, 
as projected onto the `optical plane' of a distant observer, is formed by photons freely moving along geodesics of the black hole's spacetime.  
In principle then, a shadow image like that of M87* is a \emph{geodetic} `experiment' that could enable tests of GR via the so-called `Kerr hypothesis', that is, the 
theoretically predicted uniqueness of the Kerr metric as the correct description of astrophysical  black holes. 
As far as Kerr black holes are concerned, and assuming a source of illumination of angular size much greater than the hole itself,  it is known that light rays that are 
asymptotically captured at the location of the unstable photon orbit give rise to a shadow that is nearly circular-shaped provided the black hole spin is not close to the 
maximum allowed limit~\cite{bardeen73, takahashi04}. This `Bardeen shadow', which also coincides with the black hole's capture cross section for light rays incoming 
from infinity and moving parallel to the equatorial plane, can de defined in a mathematically invariant way by expressing its equatorial and polar radii in terms of the 
corresponding impact parameters which themselves are combinations of the geodesic constants of motion~\cite{bardeen73}.

These considerations were exploited in a recent EHT paper~\cite{Psaltis_etal2020} where the shape of the M87* shadow was used to set limits on the
non-GR parameters of Johannsen's deformed Kerr metric~\cite{Johannsen:2013pca} 
%
%
(see, however, Refs.~\cite{kggp21,voelkel21} for a discussion on the limitations of black hole shadows as probes of GR gravity, when the deviations 
arise purely from non-GR gravitational degrees of freedom and are completely unrelated to accretion physics~\cite{ozel22, younsi23}).

A closely related notion has to do with the connection between the shape of a black hole shadow and the hole's mass and spin \emph{multipole moments} 
$\{ M_\ell, S_\ell \}$ (where the index $\ell$ is a non-negative integer).  According to GR's no-hair theorem, Kerr black holes are characterised by finely tuned set of multipole 
moments that are fully determined algebraically by the first two, the mass $M=M_0$ and spin angular momentum $J=S_1$. The theorem is encapsulated in the
formula~\cite{hansen74},
\be
M_\ell + i S_\ell = M ( i J/M )^\ell.
\label{nohair}
\ee
With the help of a Kerr-like metric with an adjustable quadrupole moment $M_2$, previous work has shown that a deviation from the Kerr quadrupole moment $M_2 = -J^2/M$ 
manifests itself as an oblate or prolate deformation of the Kerr shadow~\cite{Johannsen2010ApJ, Broderick2014ApJ}. These results have led to the commonly stated claim that 
the no-hair theorem itself is testable by the shape of a black hole shadow.

A nearly circular shadow is not an exclusive characteristic of Kerr black holes; in fact there are known non-Kerr black hole spacetimes that enjoy the same property. 
Examples include the aforementioned Johannsen metric~\cite{Johannsen:2013pca, Johannsen2013ApJ} and the more general metric of Carson \& Yagi~\cite{Carson_Yagi2020}. 
These spacetimes, however, are special in the sense that they are separable, i.e. they admit a third integral of motion for geodesic motion (in Kerr this is the 
well-known Carter constant~\cite{MTW1973}). This observation could be taken as evidence for  a plausible intrinsic relation between the near circularity of the shadow 
and the spacetime's separability.

In this paper, we make contact with the above considerations and assess to what extent a shadow image like that of M87*
could provide a reliable test of the no-hair theorem relation~\eqref{nohair}. To this end, we revisit the dependence of the shadow shape 
on the multipole moments by making use of stationary-axisymmetric spacetime metrics with an \emph{arbitrary} structure in the quadrupole and higher 
multipole moments. 

Our analysis is performed within the framework of GR and the employed metrics are vacuum solutions of the theory, taking the form of an expansion 
in the spin $J$ or in the inverse radial distance $1/r$ and the multipole order. It should be emphasised that, as a consequence of GR's 
uniqueness theorems for black holes~\cite{MTW1973}, these metrics do not represent true black holes in the sense that they are infested with 
horizon-piercing curvature singularities. Nevertheless, as long as the spacetime's circular photon orbit does not approach the event horizon,
this pathology has little bearing on the properties of the shadow. Alternatively, the central body could have a material surface (with negligible emission) 
instead of a horizon, i.e. something akin to a `gravastar'~\cite{mazur04}. In both cases the aforementioned spacetimes can describe the exterior 
spacetime of the putative non-Kerr compact object or even serve as proxies for the spacetime of non-GR black holes. These are precisely the 
systems likely to violate the no-hair theorem of canonical Kerr black holes. As a secondary topic of our paper we provide an analysis of the multipolar 
structure of the Johannsen metric in order to explore a connection (if any) to the spacetime's Kerr-like shadow. 

A key assumption underpinning our analysis is that the deformation away from Kerr, and the ensuing `decircularisation' of the shadow, is caused by the rotation 
of the  `black hole'. As a consequence, the effect is \emph{maximised} at the equatorial plane where the shadow radius of a backlit system is identical to the impact 
parameter $\bph$ associated with the radius $\rph$ of the unstable equatorial photon orbit (the so-called light ring). This implies that as long as we are limited 
to modest deviations from the Kerr spacetime, the photon orbit impact parameter provides an accurate measure of the shadow's non-circularity (as, in fact, it does for 
Kerr black holes).

The remainder of the paper is organised as follows. In Section~\ref{sec:axistationary} we describe the formalism for the calculation of the 
radius and associated impact parameter of the equatorial light ring of a general stationary-axisymmetric spacetime. The following two sections
comprise the paper's main calculation and results. In Section~\ref{sec:HTshadow} we obtain relations for the shadow equatorial radius 
as a function of the first few multipole moments in the Hartle-Thorne vacuum spacetime. A similar calculation is repeated in Section~ \ref{sec:axisym} 
in the context of another general stationary-axisymmetric spacetime. Section~\ref{sec:Jmetric} is dedicated to the multipole
moment analysis of the Johanssen metric. Our concluding remarks can be found in Section~\ref{sec:conclusions}. 
Throughout the paper we adopt relativistic units $G=c=1$ and use a prime to denote a radial derivative.


\section{Photon ring and impact parameter: general formalism}
\label{sec:axistationary}

The formation of a shadow is the manifestation of extreme light bending in the spacetime of a massive `dark' body,
caused by the presence of an unstable photon orbit, the so-called `light ring'. Therefore, the first step of our analysis 
is the calculation of the light ring radius $\rph$ and the associated impact parameter $\bph$ which describes photons that approach 
the black hole from infinite distance and get trapped at the light ring.

Here we consider an arbitrary axisymmetric and stationary metric of the form
\be
ds^2 = g_{tt} dt^2 + g_{rr} dr^2 + 2g_{t\varphi} dt d\varphi + g_{\theta\theta} d\theta^2 + g_{\varphi\varphi} d\varphi^2,
\label{metric}
\ee
with $ g_{\alpha\beta} = g_{\alpha\beta} (r,\theta)$, assuming a spherical-like coordinate system. 
The assumed symmetries allow us to write the following equations for the $u^t, u^\varphi$  four-velocity components,
\be
u^t = \frac{1}{\cD} \left (\,  g_{t\varphi} b + g_{\varphi\varphi}  \, \right ), \quad
u^\varphi = -\frac{1}{\cD} \left (\,  g_{t\varphi}  + g_{tt} b \, \right ),
\ee
where $  \cD = g_{t\varphi}^2  - g_{tt} g_{\varphi\varphi} $. The orbital constants $E$ (energy per unit mass) and 
$L$ (angular momentum per unit mass) enter through the impact parameter $b = L/E$. Upon inserting these in the 
normalisation condition $u^\mu u_\mu =0$ and setting $u^\theta=0$ we end up with a radial motion equation,
\be
 g_{rr} (u^r )^2  = \frac{1}{\cD} \left (\, g_{tt} b^2 + 2 g_{t\varphi} b + g_{\varphi\varphi} \, \right ) \equiv   V_{\rm eff} (r ,b).
\ee
A circular orbit obeys 
\be
 V_{\rm eff} ( \rph ,\bph) = 0, \qquad   V_{\rm eff}^\p (\rph ,\bph) = 0.
\ee
These two conditions lead to the light ring equation,
\begin{align}
& \Big \{ 4 \left (  g_{t\varphi}  g_{tt}^\p   -  g_{tt}  g_{t\varphi}^\p  \right )   
\left (  g_{t\varphi} g_{\varphi\varphi}^\p    - g_{\varphi\varphi} g_{t\varphi}^\p +  \right ) 
\nonumber \\
&  + \left ( g_{\varphi\varphi} g_{tt}^\p - g_{tt} g_{\varphi\varphi} ^\p \right )^2 \Big \}_{r=\rph} =0,
\label{photonEq}
\end{align}
and the associated impact parameter,
\be
\bph= \frac{1}{2} \frac{ g_{\varphi\varphi} g_{tt}^\p - g_{tt} g_{\varphi\varphi} ^\p}{g_{tt}  g_{t\varphi}^\p -g_{t\varphi}  g_{tt}^\p }.
\label{bgeneral}
\ee 
These formulae describe both prograde and retrograde photon motion, the two cases being distinguished by the sign of the spin parameter
(which does not appear explicitly here). As a benchmark example we may consider the standard Kerr metric in Boyer-Lindquist parameters. We find,
\begin{align}
& \rph ( \rph  -3M)^2 = 4 a^2 M^3, 
\\
& \bph = M \left [ a + \left ( \frac{\rph }{M} \right )^{3/2} \right ],
\label{bphKexact}
\end{align}
where $a = J/M^2$ is the dimensionless spin parameter. Prograde (retrograde) motion would correspond to $a>0$  $(a<0)$.

From its definition (and its association with the light ring), $\bph$ coincides with the equatorial radius of the shadow cast
by a black hole that is backlit from a source of large angular size (e.g., a distant luminous plane) and viewed from the opposite direction 
by an equatorial observer. As discussed, for example, in Ref.~\cite{bardeen73}, this is a  coordinate \emph{invariant} identification. 
Rotation causes the bifurcation of the light ring into prograde (co-rotating) and retrograde (counter-rotating) branches and as a consequence
the shadow develops a left-right asymmetry.
The two orbits vary asymmetrically with the black hole spin, resulting in a relative displacement between the centre of the black hole (at $r=0$) and 
that of the non-circular shadow.  Nevertheless, knowledge of the prograde and retrograde impact parameters $ \bph = \{ b_{\rm pro}, b_{\rm retro} \}$ 
allows us to calculate an invariant equatorial shadow radius (as viewed by an equatorial observer) defined by the geometric average,
\be
\bar{b}_{\rm ph} = \frac{1}{2} \left ( b_{\rm pro} + b_{\rm retro} \right ).
\label{bphaver}
\ee
The equatorial plane (or, more generally, the latitude slice  $\theta \sim \pi/2$) is where the shadow's shape is expected to show
the most pronounced deviation from circularity because of the maximum differential dependence of $\rph$ and $ b_{\rm pro},  b_{\rm retro} $
on the spin.  The opposite arrangement is expected to happen near the symmetry axis $\theta = \{ 0, \pi \}$ where the impact
of rotation is at its minimum.  
This situation is exemplified by the shadow shape found in a number of deformed Kerr spacetimes 
(see e.g.~\cite{Johannsen2013ApJ, Carson_Yagi2020}) and of course by the Kerr shadow itself, where as the spin parameter $a \to 1$ it 
acquires a characteristic `D' shape as a result of the rapid  shrinkage of $b_{\rm pro}$ with respect to the much slower changing $b_{\rm retro}$, 
see, e.g.,  Ref.~\cite{Johannsen2010ApJ}. The example most relevant to this paper is the shadow obtained in Ref.~\cite{kostaros22}
\footnote{In that work the Hartle-Thorne metric is used `as it is', without further expansion of the geodesic equations with respect
to the spin. An interesting consequence of this approach is the emergence of non-equatorial light rings, a property not present in our model
where all equations are consistently spin expanded.}  
using the Hartle-Thorne spacetime (the subject of the next section); it is found that, in spite of the presence of moderate rotation, 
the shadow's polar radius remains very close to the radius of a Schwarzschild black hole.


\section{Shadow radius in the Hartle-Thorne spacetime} 
\label{sec:HTshadow}

\subsection{The ${\cal O} (J^4)$ Hartle-Thorne spacetime}

The purpose of this and the following section is to put on a quantitative basis the relation between the black hole shadow
size and a given spacetime's multipole moment structure. Perhaps the best available tool for undertaking this task in
GR is the celebrated \emph{exterior} Hartle-Thorne (HT) metric~\cite{HT68} in its modern incarnation of ${\cal O} (J^4)$ 
precision~\cite{Yagi_etal:2014}, where $J$ represents the central body's angular momentum.

The HT metric has the advantage of allowing arbitrary values for the first five mass and spin multipole moments, $\{M, S_1, M_2, S_3, M_4 \}$;
these are Geroch-Hansen moments, see Ref.~\cite{quevedo90} for a review. In addition, it is generically non-separable with respect to the Hamilton-Jacobi 
equation for geodesic motion, that is, it does not admit a Carter constant.  As already mentioned in the Introduction, we employ the HT metric to describe the
exterior spacetime of a rotating ultracompact body with a surface or as an effective black hole spacetime, ignoring the fact that it is not well behaved in the 
vicinity of the event horizon. This is not a problem as long as the light ring stays well clear of the horizon radius. It should be pointed out that the HT metric 
includes the Kerr spacetime as a special limit (and as an expansion in the spin) with the added property of being singularity-free (except at the center) and separable. 

The HT metric is an expansion in the spin and takes the following functional form~\cite{Yagi_etal:2014}
(here $\epsilon$ is the spin-order bookkeeping parameter, $\mu =\cos\theta$ and $\nu = \{t, r,\mu,\varphi \}$):
\begin{align}
g_{\nu\nu} (r,\mu) &= g_{\nu\nu}^{(0)} (r) + \left[ g_{\nu\nu}^{(20)} (r) + g_{\nu\nu}^{(22)} (r) P_2 (\mu) \right] \epsilon^2 
 \nonumber \\
&+ \left[ g_{\nu\nu}^{(40)} (r)  + g_{\nu\nu}^{(42)} (r) P_2 (\mu) + g_{\nu\nu}^{(44)} (r) P_4 (\mu) \right]\epsilon ^4 
\nonumber \\
& + {\cal O} (\epsilon^6 ),
\\
\nonumber \\
g_{t \varphi} (r,\mu) & = (1-\mu^2 ) \left \{ g_{t\varphi}^{(1)} (r)   \epsilon  \right.
\nonumber \\
& \left. +\left[ g_{t \varphi}^{(31)} (r) + g_{t\varphi}^{(33)} (r) \frac{d P_3}{d\mu} (\mu) \right]\epsilon^3 \right \} 
+ {\cal O} (\epsilon^5 ),
\end{align}
where $P_n (\mu)$ is the standard Legendre polynomial. The coordinate change $\mu \to \theta$ in the line element results in 
$ g_{\theta \theta} (r,\theta) =  g_{\mu \mu } \sin ^2 \theta$. The radial metric functions $g_{\mu\nu}^{(nm)}(r)$ are given by rather lengthy expressions 
and therefore are not presented here. The interested reader can find them in Ref.~\cite{Yagi_etal:2014}.

Apart from their dependence on the coordinates $\{r,\mu\}$, the metric components depend on a number of parameters 
associated with the central body. These are
\be
\{ M_\rS, J, C_{20}, C_{40}, C_{22}, C_{42}, C_{44}, C_{31}, C_{33} \},
\ee
where $M_\rS$ is the mass of the spherical body in the limit of zero rotation. A given $C_{nm}$ parameter is of order $\epsilon^n$
in the spin and appears in tandem with the $P_m (\mu)$ polynomial. As we are about to see, these are related to the source's first five
multipole moments, $\{M, S_1, M_2, S_3, M_4 \}$. The simplest way for extracting the HT moments is to expand $g_{tt}, g_{t\varphi}$ in powers of $1/r$,
\begin{widetext}
\begin{align}
g_{tt} & = -1 + \frac{2 M_\rS}{r} ( 1 + C_{20} \epsilon^2 + C_{40} \epsilon^4) - \frac{2}{r^3} \left \{ C_{42} M_\rS^3 \epsilon^4
+ \left ( \frac{J^2}{M_\rS} + \frac{8}{5} C_{22} M^3_\rS \right ) \epsilon^2  \right \} P_2 
\nonumber \\
&  + \frac{2 P_4}{r^5} \left [ \frac{107}{105} \frac{J^4}{M_\rS^3} + \frac{9428}{735} C_{22} J^2 M_\rS - \frac{4}{7} C_{33} J M^3_\rS
+ \left (  C_{44} + \frac{3144}{245} C_{22}^2 \right ) M_\rS^5  \right ] \epsilon^4  + {\cal O} \left (\frac{\epsilon^2}{r^4}, \frac{1}{r^6}, \epsilon^6 \right ),
\label{gexpand1}
\\
\nonumber \\
g_{t\varphi} & = \sin^2\theta \left \{  -\frac{2}{r} \left ( J \epsilon +  C_{31} M_\rS^2 \epsilon^3 \right ) 
- \frac{2}{3} \frac{C_{33} M_\rS^4}{r^3} \frac{d P_3}{d\mu}   \epsilon^3 \right \}  + {\cal O} \left ( \frac{1}{r^5}, \epsilon^5 \right ).
\label{gexpand2}
\end{align}
\end{widetext}
Following the same practice as in Newtonian gravity, we can read off the multipoles from the coefficients of  the $1/r$ powers (see discussion in Ref.~\cite{Pappas:2012}).
 We find,
\begin{align}
M &= M_0 =  M_\rS \left ( 1 + C_{20} \epsilon^2 + C_{40} \epsilon^4 \right )  + {\cal O} \left (\epsilon^6 \right ), 
\\
\nonumber \\
S_1 &=   J \epsilon +  C_{31} M_\rS^2 \epsilon^3  + {\cal O} \left (\epsilon^5 \right ),  
\\
\nonumber \\
M_2 & =  -\left ( \frac{J^2}{M_\rS} + \frac{8}{5} C_{22} M^3_\rS \right ) \epsilon^2 -  C_{42} M_\rS^3 \epsilon^4 + {\cal O} \left (\epsilon^6 \right ), 
\\
\nonumber \\
S_3 & = C_{33} M_\rS^4 \epsilon^3  + {\cal O} \left (\epsilon^5 \right ), 
\\
\nonumber \\
M_4 & =  \left [ \frac{107}{105} \frac{J^4}{M_\rS^3} + \frac{9428}{735} C_{22} J^2 M_\rS - \frac{4}{7} C_{33} J M^3_\rS
\right.
\nonumber \\
&\left. \quad + \left (  C_{44} + \frac{3144}{245} C_{22}^2 \right ) M_\rS^5 \right ] \epsilon^4 + {\cal O} \left (\epsilon^6 \right ).
\end{align}
The mass correction terms can be set to zero, $C_{20} = C_{40} = 0$, so that we can replace  $M_\rS \to M$ in all expressions henceforth. 
This amounts to a simple shift in the mass scale of the system without any further physical importance.

The Kerr limit of the HT spacetime is retrieved for
\begin{align}
C_{22}  &= C_{42} = C_{31} = 0, 
\\
C_{33} &= - \frac{J^3}{M_\rS^6}, ~C_{44}= - \frac{62}{105} \frac{J^4}{M^8_\rS},
\end{align}
and corresponds to the following set of multipole moments [see Eq.~\eqref{nohair}]: 
\be
 S_1^\rK = J, ~M_2^\rK = - \frac{J^2}{M}, ~S_3^\rK = - \frac{J^3}{M^2}, ~M_4^\rK = \frac{J^4}{M^3}.
\ee
(In this and other formulae below, the index `K'  labels a Kerr metric quantity.)

Instead of the original $C_{nm}$ parameters, it is more convenient to work with a new set of dimensionless  
parameters $c_{nm}$, defined as
\be
C_{nm} = c_{nm} \frac{J^n}{M^{2n}}.
\ee
We then define the Kerr \emph{deviation} parameters $\varepsilon_{nm}$ as
\be
c_{nm} = c_{nm}^\rK + \varepsilon_{nm}, 
\label{Kerrdeviation_params}
\ee
with $\left (  c_{22}^\rK,  c_{42}^\rK ,  c_{31}^\rK ,  c_{33}^\rK,  c_{44}^\rK \right ) = ( 0, 0, 0, -1, - 62/105) $.


\subsection{The equatorial shadow radius}

\begin{figure}[htb!]
\begin{center}
\includegraphics[width=\columnwidth]{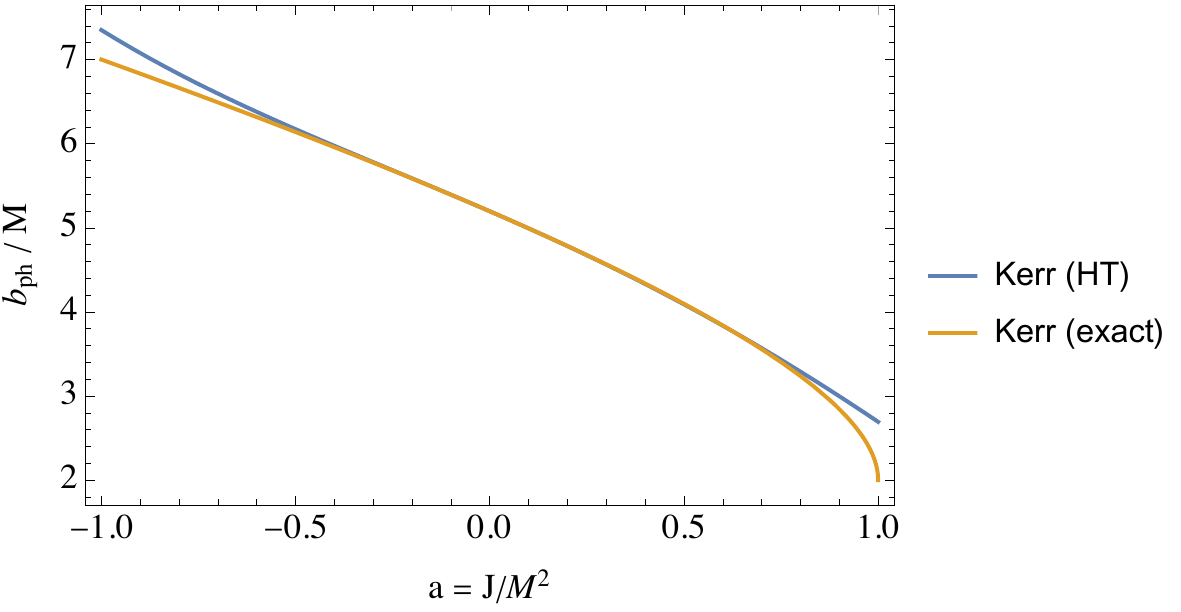}
\end{center}
\caption{Comparison of the light ring impact parameter $\bph$, calculated using the Kerr limit of the HT spacetime, Eq.~\eqref{bphHT},
and the exact Kerr metric, Eq.~\eqref{bphKexact}, across the allowed spin range $-1 \leq  a \leq 1$. The prograde (retrograde) $\bph$ 
corresponds to $a >0$ ($a<0$).}
\label{fig:KerrTest}
\end{figure}

It is straightforward to use the general expressions~\eqref{photonEq} and \eqref{bgeneral} in the HT spacetime and obtain the light ring radius $\rph$ and 
the associated impact parameter  $\bph $ as expansions in the dimensionless spin parameter $a = J/M^2$ (or in $\epsilon$). For example, for the latter parameter 
we find
\begin{widetext}
\begin{align}
\frac{\bph}{M} &= 3\sqrt{3} - 2 a + \frac{1}{2\sqrt{3}} \left [ -1 + 18 c_{20} + 9 c_{22} (-16 + 15\log3) \right ] a^2  
 \nn \\
 & + \left  [ \frac{425957471}{15552} + \frac{605}{8} c_{20} + \frac{27}{2} c_{40} + \frac{585}{8} c_{42} + \frac{1534869}{128} c_{44} 
 - \frac{6380937}{256} \log 3 - \frac{2025}{32} \log 3\, c_{20}  - \frac{2025}{32} \log 3 \, c_{42} \right.
 \nn \\
& \left.   - \frac{5587785}{512} \log 3\, c_{44}  + \frac{9}{2}  ( 45 \log 3 -52) c_{20} c_{22} + \frac{c_{31}}{48} ( 6075 \log 3 -7132) 
- \frac{c_{33}}{384}  ( 7096815 \log 3 -7798684 ) \right.
\nn \\
& \left. - \frac{13}{40320}  ( 510678081\log 3 -561143860  ) c_{22} + \frac{3}{1120}  \left (41913428 - 39495033 \log 3 + 1225854 \log 3^2 \right ) c_{22}^2
\right ] a^3 
\nn \\
& + \frac{a^4}{19595520 \sqrt{3}} \Big [ -176359680 c_{20}^2 - 11664 \left ( 2520553852 - 3237502203 \log 3 + 858538926 \log 3^2 \right )   c_{22}^2
\nn \\
& + 20412 c_{20} \left \{ 8417925 \log 3 -9265540 + 8640 c_{31}   +   ( 8037225 \log 3 - 8848980 )  c_{33} + 48 c_{22} ( 500175 \log 3 -550652) \right \} 
\nn \\
& - 7 \left \{ 588719779672 + 50388480 c_{40} + 1214222400 c_{42} + 272213921700 c_{44} - 535888505070 \log 3 
\right. 
\nn \\
&  \left. - 1110121200  \log 3\, c_{42} - 247788794925 \log 3\, c_{44}  + 38880  ( 57105 \log 3 -62576) c_{31} - 16200 ( 24047847 \log 3 -26418796)  c_{33} \right \}
\nn \\
&  + 108 c_{22} \left \{ -292460091548 + 296252520363 \log 3 - 27342639450 \log3^2 + 181440 c_{31} ( 297 \log 3 -310 ) 
\right.
\nn \\
&  \left. - 39690 c_{33} \left (842680 - 1523826 \log 3 + 688905 \log3^2 \right ) \right \} \Big ] + {\cal O} (a^5).
\label{bphHT}
\end{align}
\end{widetext}
With the help of the Kerr limit of this expression we can benchmark the precision of the HT metric if we compare it against the full Kerr result~\eqref{bphKexact}. 
Although these formulae assume different coordinate systems, the comparison is nevertheless meaningful because $\bph$ is itself a gauge-invariant
quantity. The outcome of this exercise is shown in Fig.~\ref{fig:KerrTest} where we plot $\bph (a)$.
It is clear that, at least as far as $\bph$ is concerned, the ${\cal O} (J^4 )$ HT metric performs extremely well for
 $ -0.8 \lesssim a \lesssim 0.8 $. We expect the same level of accuracy to be representative of the 
 HT spacetime in general.

\begin{figure}[htb!]
\begin{center}
\includegraphics[width=\columnwidth]{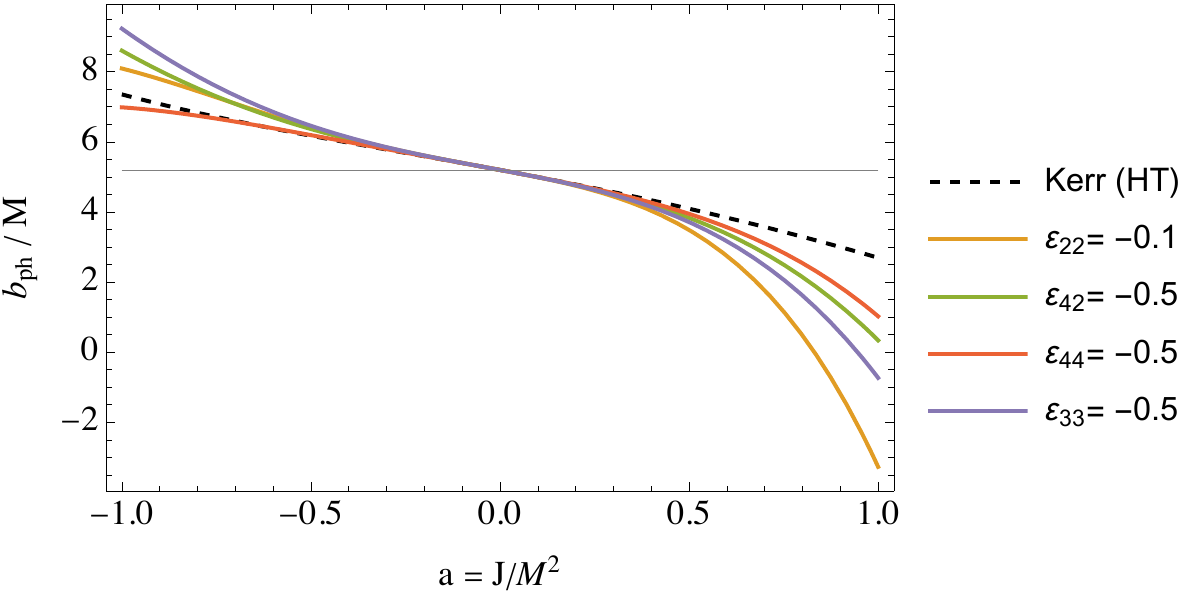}
\\ \vspace{0.5cm}
\includegraphics[width=\columnwidth]{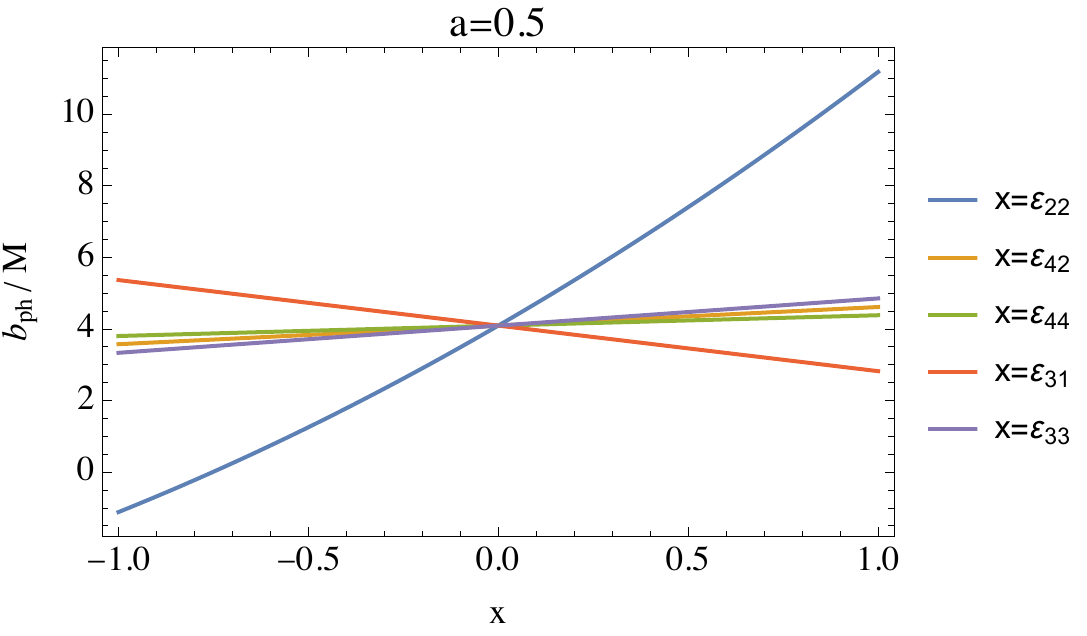}
\end{center}
\caption{The light ring impact parameter $\bph$ of the HT spacetime as a function of the dimensionless spin $a=J/M^2$
and the Kerr deviation parameters $\varepsilon_{nm}$ defined in Eq.~\eqref{Kerrdeviation_params}. Top panel: 
individually fixed $\varepsilon_{nm} <0$ and varying $a$ (both prograde $a>0$ and retrograde $a<0$ cases). 
The plot also shows the Schwarzschild impact parameter $b_{\rm Schw}$ (thin horizontal line) as well as the Kerr limit of the HT $\bph$ (dashed curve).
Bottom panel: fixed spin $a=0.5$ and individually varying deviation parameters $ x=\varepsilon_{nm}$.}
\label{fig:bphHT}
\end{figure}

Moving away from Kerr, in Fig.~\ref{fig:bphHT} we show $\bph$ as a function of $a$ for fixed $ \varepsilon_{nm}$
(top panel) and as a function of the deviation parameters for $a = 0.5$  (bottom panel). 
According to the displayed results, a moderate deviation $ | \varepsilon_{nm} | \sim 1 $ from Kerr is more than
enough to cause a notable change in $\bph$, especially for the prograde case, even for a moderate value of the spin. 
As expected from the expansions~\eqref{gexpand1} and \eqref{gexpand2}, the quadrupole parameter $\varepsilon_{22}$ is the one 
typically associated with the largest deviation. 

\begin{figure}[htb!]
\begin{center}
\includegraphics[width=\columnwidth]{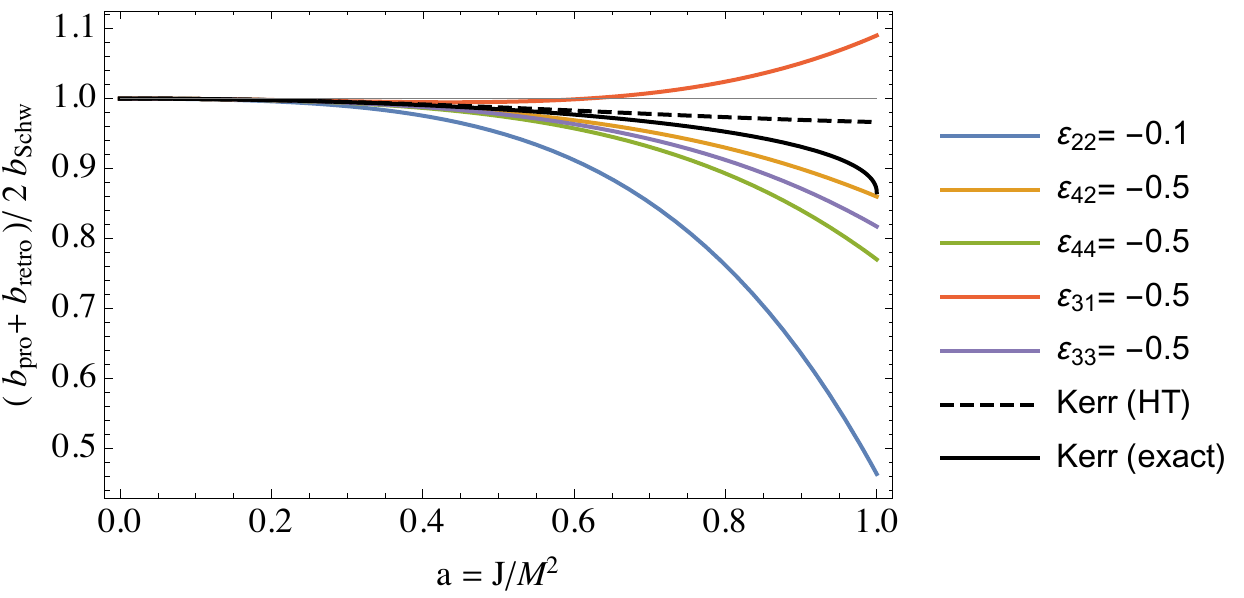}
\\ \vspace{0.2cm}
\includegraphics[width=\columnwidth]{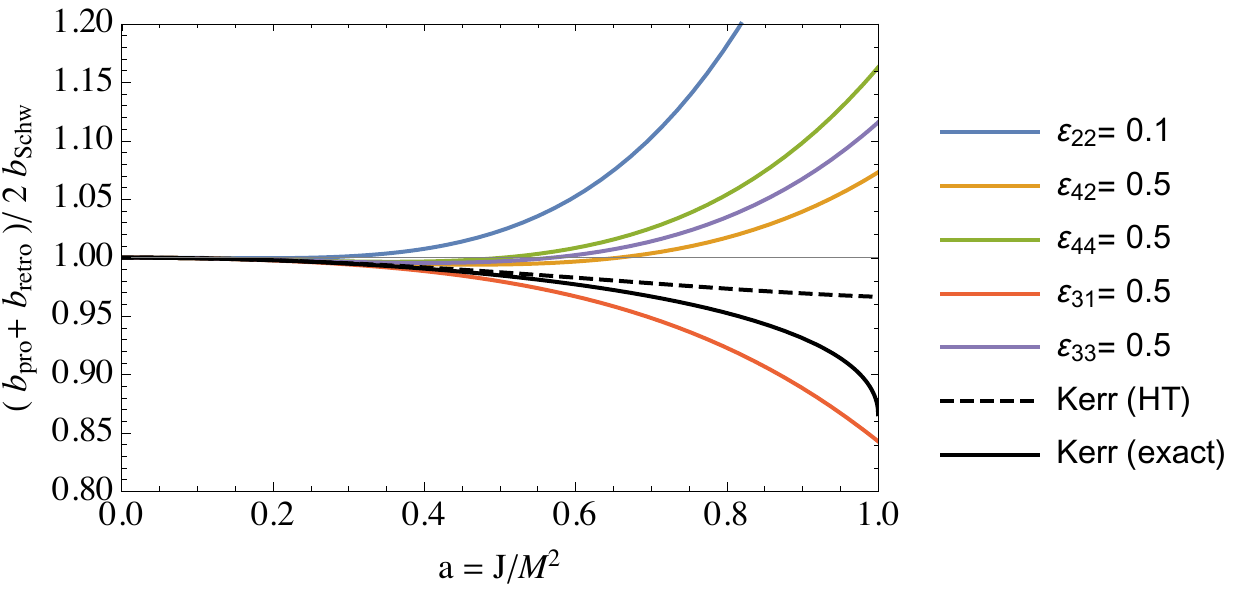}
\end{center}
\caption{The HT spacetime's equatorial shadow radius $\bar{b}_{\rm ph} = (b_{\rm pro} + b_{\rm retro} )/2$ (normalised to the Schwarzschild 
radius $b_{\rm Schw}$) as a function of the spin $a$ for $\varepsilon_{nm} <0$ and  $\varepsilon_{nm} > 0$ Kerr
deviation parameters (top and bottom, respectively). In all cases only a specific $\varepsilon_{nm}$ is set to a non-zero value. 
For the purpose of comparison the figure includes the radius  $\bar{b}_{\rm ph}$ for the exact Kerr metric (solid black curves) and for
the Kerr limit of the HT metric (dashed curves).}
\label{fig:meanbHT1}
\end{figure}

The key parameter in relation with the shape of the shadow is the averaged prograde-retrograde impact parameter $\bar{b}_{\rm ph}$
as defined in Eq.~\eqref{bphaver}. This parameter is obtained most easily by simply removing the odd-order spin terms in an expansion like 
the one in Eq.~\eqref{bphHT}; these are the terms that cancel out when we sum $b_{\rm pro}$ and $b_{\rm retro}$. 
 As already discussed, $\bar{b}_{\rm ph}$ coincides with the equatorial radius of the shadow cast by a backlit black hole, as viewed by an equatorial observer.  
This is displayed in Fig.~\ref{fig:meanbHT1} in the form  $\bar{b}_{\rm ph}/b_{\rm Schw}$ (where $ b_{\rm Schw} = 3 \sqrt{3} M$ is the radius of the Schwarzschild 
black hole's circular shadow) as a function of $a$ 
and for different values of the deviation parameters $ \varepsilon_{nm} $ (each one `switched on' individually). The top (bottom) panel shows 
results for $\varepsilon_{nm} >0$ ( $\varepsilon_{nm} < 0$); in general, flipping sign in a given $\varepsilon_{nm}$ causes $\bar{b}_{\rm ph}$ 
to move to the opposite side of the circular $b_{\rm Schw}$ radius. All $\varepsilon_{nm}$ parameters appear to be correlated in this respect, 
with the exception of $\varepsilon_{31}$ which is anticorrelated. 

The results of Fig.~\ref{fig:meanbHT1} suggest that a moderate $ | \varepsilon_{nm} | \sim 0.5$ deviation could produce a minimum $ 10\%-20 \%$ 
variation with respect to a circular shadow for $ 0.6 \lesssim a \lesssim 0.8$. At the same time the Kerr `yardstick' discussed earlier suggests 
that the accuracy of the HT shadow radius begins to deteriorate at  $a \approx 0.8$. 
Once again, the quadrupole deformation $\varepsilon_{22}$ stands out as the shadow's dominant decircularisation factor. 
The moderate degree of variation of the equatorial radius with $a$ implies that the corresponding change in the shadow's \emph{polar} radius 
should be even less pronounced, thus essentially retaining its zero-rotation value $ b_{\rm Schw}$.

\begin{figure}[htb!]
\begin{center}
\includegraphics[width=\columnwidth]{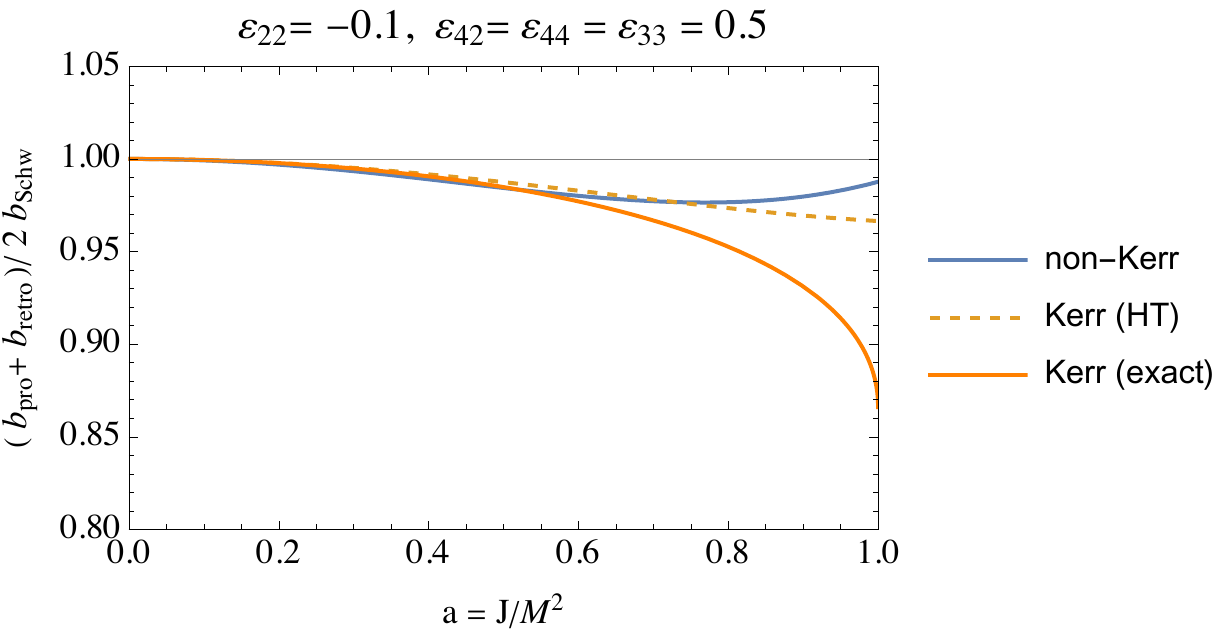}
\end{center}
\caption{An example of a HT spacetime with a nearly circular shadow despite a strongly non-Kerr structure.  We show the
normalised equatorial shadow radius $\bar{b}_{\rm ph}/b_{\rm Schw}$ as a function of the spin $a$ for Kerr deviation parameters
$ ( \varepsilon_{22},  \varepsilon_{33}, \varepsilon_{42}, \varepsilon_{44} ) = (-0.1, 0.5, 0.5, 0.5) $.  The HT result (designated as `non-Kerr') 
is compared against the shadow radius of the full Kerr spacetime (solid curve) and the Kerr limit of the HT spacetime (dashed curve).
}
\label{fig:meanbHT2}
\end{figure}

A perhaps surprising situation could arise when several of the deformation parameters are present at the same time. 
For instance, it is fairly easy to produce a very Kerr-like shadow that remains nearly circular for the entire spin range  
despite having a strongly non-Kerr multipolar structure. An example is shown in Fig.~\ref{fig:meanbHT2} where a 
dominant $\varepsilon_{22} < 0$ is counterbalanced by the combined presence of  $\varepsilon_{42}, \varepsilon_{44}, \varepsilon_{33} >0$.
What this really means is that the connection between the shadow shape and the multipole moments of the spacetime (or their proxies 
$\varepsilon_{nm}$) is not as direct as suggested in previous work \cite{Johannsen2010ApJ, Broderick2014ApJ}.  This important issue is further 
explored in the following section where we express $\bph$  and the shadow in terms of the multipole moments themselves. 


\subsection{The shadow radius as a function of the multipole moments  $M_\ell, S_\ell$.}
\label{sec:bphMM}

The previous section provided us with some understanding of the dependence of the equatorial 
shadow radius as a function of the original $C_{nm}$ parameters of the HT metric, expressed as deviations from 
the Kerr spacetime. Here we complete this analysis by expressing the shadow radius directly in terms of the Geroch-Hansen 
multipole moments $\{ M_\ell, S_\ell \}$.

As a first step, we define a new set of dimensionless moments, $ \{ \cM_\ell, \cS_\ell \}$,
by factoring out the spin dependence together with the suitable powers of $M$,
\be
S_\ell = \cS_\ell \frac{J^\ell}{M^{\ell-1}}, \quad M_\ell = \cM_\ell \frac{J^\ell}{M^{\ell-1}}.
\label{MMnondim}
\ee
The next step is to express $J, C_{nm} $ in terms of these new parameters and the spin $a$. 
In particular, $J$ is replaced with the help of  $J \to S_1 - C_{31} M^2 \epsilon^2 $ which means that the parameter 
$c_{31}$ is expected to be present in $\bph$. The same is true for $c_{42}$ after replacing $ C_{22} \to C_{22} ( \cM_2, J, c_{42} )$. 

Once $\bph (a,  \cM_\ell, \cS_\ell) $ is calculated, we can easily obtain the shadow radius $\bar{b}_{\rm ph}$. The 
final ${\cal O} (a^4)$ expression is
\begin{widetext}
\begin{align}
\frac{\bar{b}_{\rm ph}}{M} &= 3\sqrt{3} + \frac{a^2}{16 \sqrt{3}} \left [\, \cS_1^2 ( 712 - 675 \log 3 )
+ 45 \cM_2 (16 - 15 \log3 )\, \right ] + \frac{a^4}{\sqrt{3}} \left [\,  \frac{5}{8}  c_{42} (-622 + 567\log 3 )
\right.
\nn \\
\nn \\
& \left. + \frac{315}{256} \cM_4 (-79028 + 71937 \log 3)   + \cS_1^4 \frac{ \left (403363330924 + 2226769271865 \log 3 - 2361245507100 \log 3^2 \right )}{22394880}
\right.
\nn \\
\nn \\
& \left. + \cM_2 \cS_1^2 \frac{\left ( 2538443668 + 25485296247 \log 3 - 25302890160 \log 3^2 \right )}{82944}
+ \cS_1 c_{31} \frac{(70118 - 64395 \log 3)}{72}
\right.
\nn \\
\nn \\
& \left. + \frac{35}{6912} \cS_1 \cS_3 \left (-18414152 - 3671082 \log 3 + 18600435 \log 3^2 \right)
+ \frac{35}{256} \frac{\cM_2 \cS_3}{\cS_1} \left (842680 - 1523826 \log 3 + 688905 \log 3^2 \right ) 
\right.
\nn \\
\nn \\
& \left. + \frac{5}{6144} \cM_2^2 \left (121182452 - 379769553 \log 3 + 245296836 \log 3^2 \right)
+ \frac{5}{8} \frac{ \cM_2 c_{31}}{\cS_1} (310 - 297 \log 3)
\, \right ].
\label{bphMM1}
\end{align}
The numerical value of this result is given by the much shorter expression:
\begin{align}
\frac{\bar{b}_{\rm ph}}{M} &\approx 5.19615  - a^2 \left ( 0.778098\, \cM_2   + 1.06677\, \cS_1^2 \right ) 
+ a^4 \Big (  0.329511\, c_{42} - 11.1427 \,\cM_2^2 + 2.04045\, \cM_4 
\nn \\
& - 16.6617\, \cM_2 \cS_1^2    - 4.72798 \, \cS_1^4  -5.87737\, c_{31}  \frac{\cM_2}{\cS_1}
+  4.67328 \,  \frac{\cM_2 \cS_3 }{\cS_1} 
-5.02887 \, c_{31}  \cS_1 + 7.39031\,   \cS_1 \cS_3   \Big ).
\label{bphMM2}
\end{align}
\end{widetext}
Let us first focus on the ${\cal O} (a^2)$ portion of this formula; we can see that 
 a $\cM_2 <0$ quadrupole counteracts the shadow shaping action of the `frame-dragging' multipole $\cS_1$.
This is indeed what happens in Kerr, which has $\cS_1^\rK =1$ and $\cM_2^\rK =-1$ and is in agreement with
the analysis of Ref.~\cite{Broderick2014ApJ}.  A near perfect cancellation of the two effects takes place  for
\be
\cM_2 \approx - 1.371 \cS_1^2,
\label{M2circHT}
\ee
which means that non-Kerr bodies too can produce quasi-circular shadows. At the other end of the spectrum,
prolate bodies have $\cM_2 > 0$ and are likely to cast a markedly non-circular shadow.  

\begin{figure}[htb!]
\begin{center}
\includegraphics[width=0.73\columnwidth]{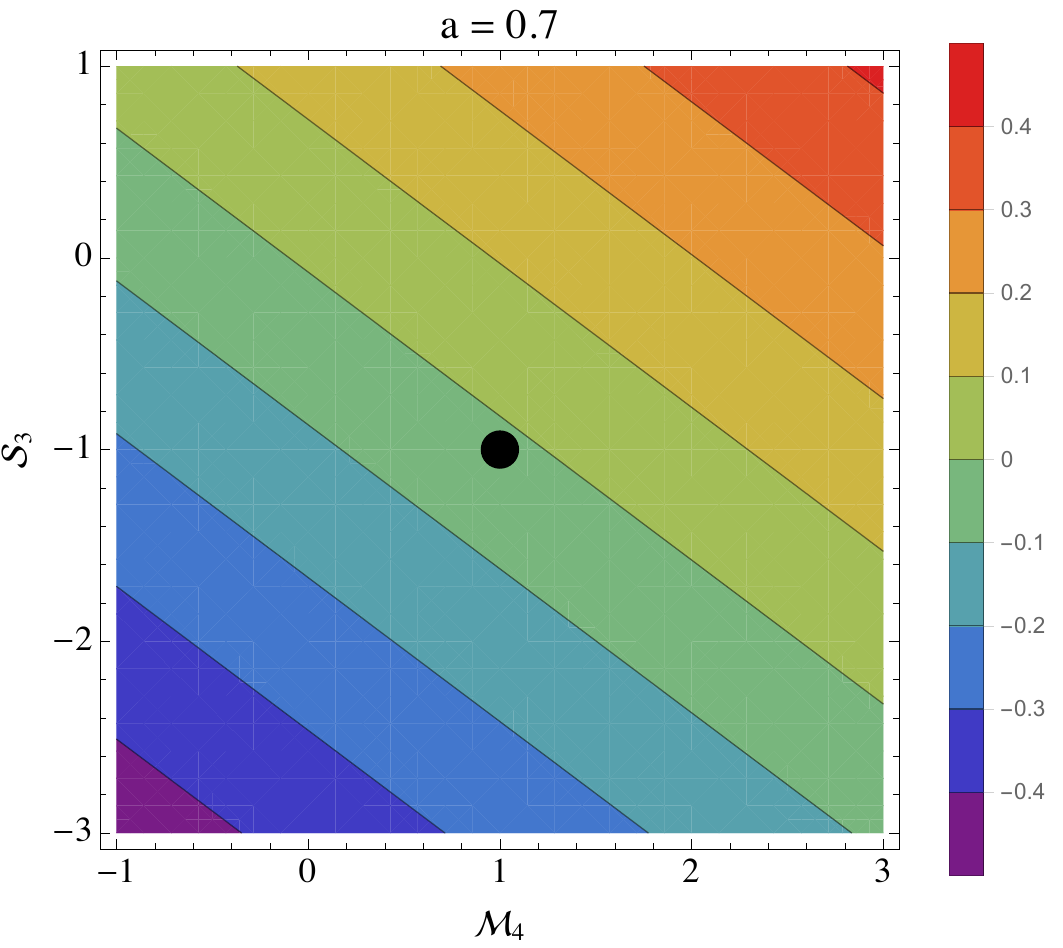}
\\ \vspace{0.1cm}
\includegraphics[width=0.73\columnwidth]{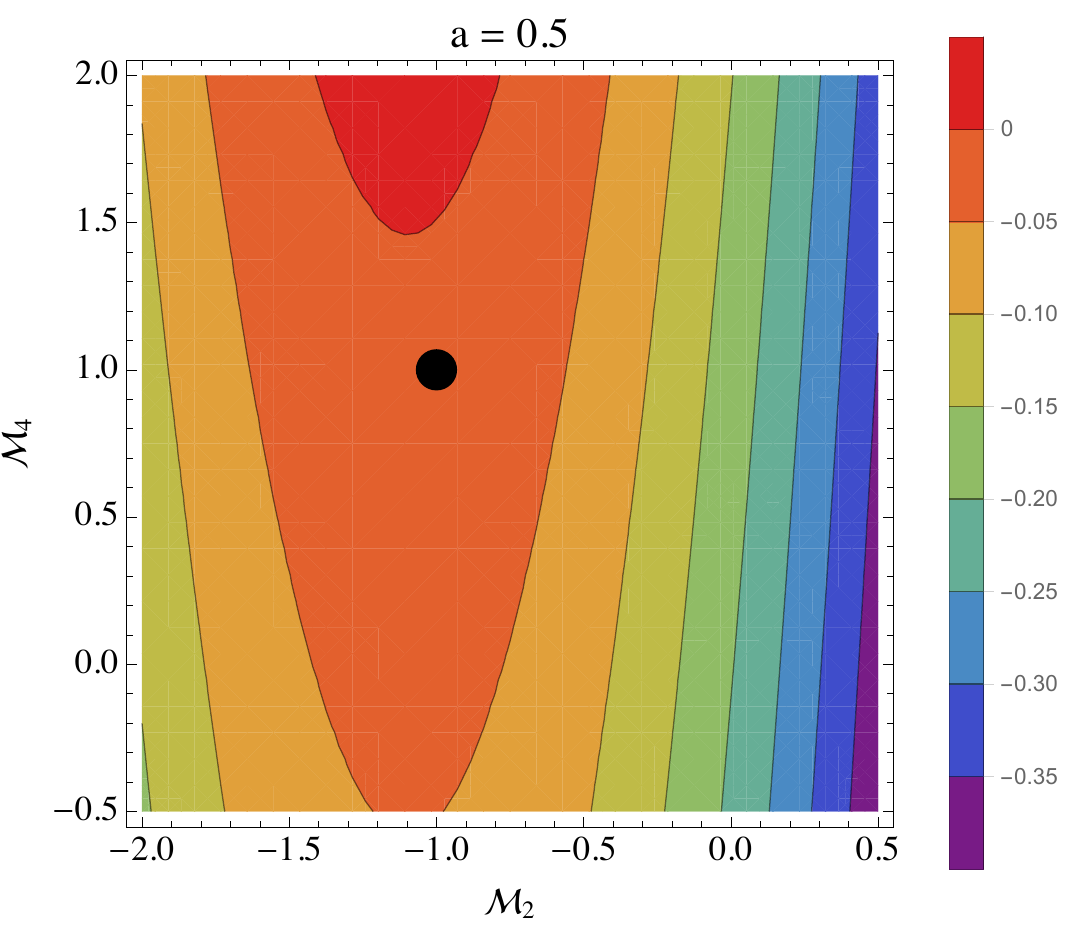}
\\ \vspace{0.1cm}
\includegraphics[width=0.73\columnwidth]{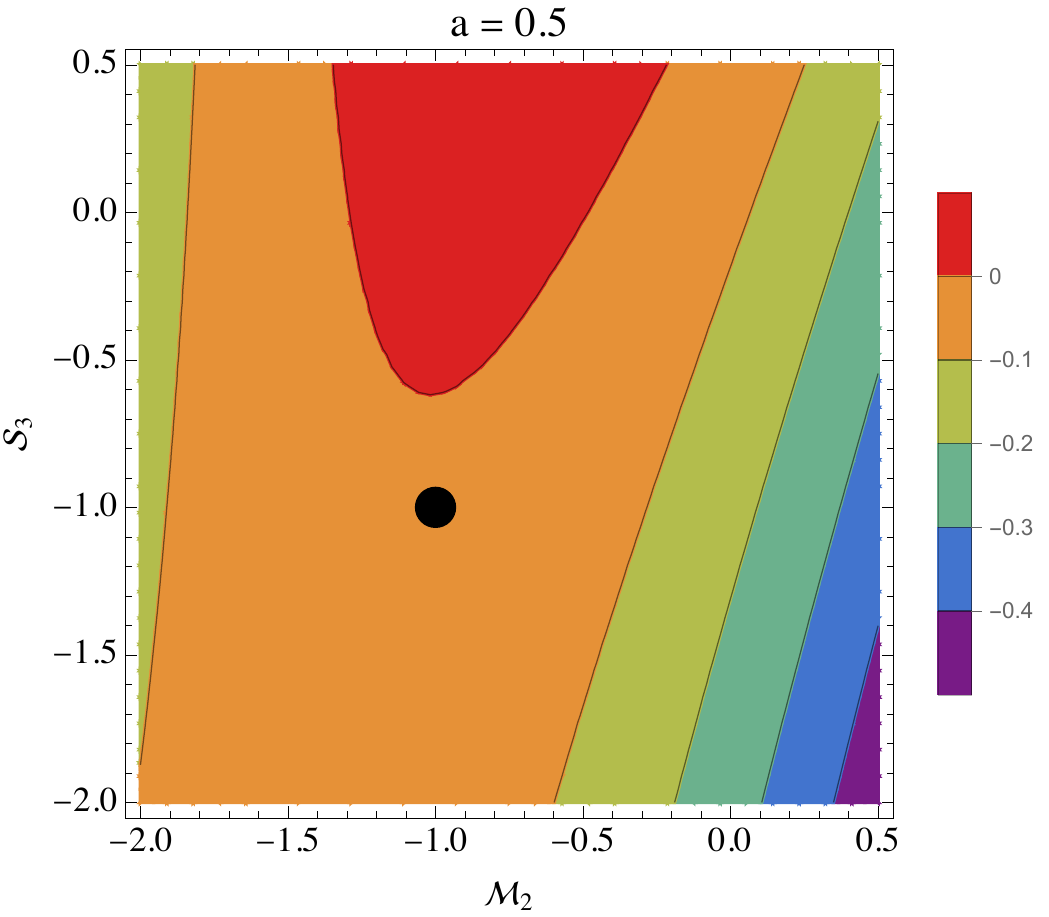}
\end{center}
\caption{Contour plot of the equatorial HT shadow radius expressed as a fractional difference 
$ \Delta_b =  \bar{b}_{\rm ph}/b_{\rm Schw} -1 $ with respect to the circular Schwarzschild radius. 
Top panel: $\Delta_b$ is shown as a function of the $\cM_4, \cS_3$ multipole moments for spin  $a=0.7$,
see Eq.~\eqref{baverM4S3}. The remaining moments have been set equal to their Kerr values, 
$ \cM_2 = -1, \cS_1 = 1$.
Middle panel: $\Delta_b$ is shown as a function of the $\cM_2, \cM_4$ mass moments for $a=0.5$,
see Eq.~\eqref{baverM2M4}. The remaining moments have been set equal to their Kerr values,  
$\cS_1 =1, \cS_3= -1$.
Bottom panel: $\Delta_b$ is shown as a function of the $\cM_2, \cM_4$ mass moments for $a=0.5$,
see Eq.~\eqref{baverM2M4}. The remaining moments  $\cS_1, \cS_3$  have been set equal to their Kerr values. 
In all cases, we have set the remaining metric constants $c_{42}, c_{31}$ equal to zero.
The black dot marks the HT Kerr  limit of $\Delta_b$.}
\label{fig:bHTcontour}
\end{figure}

The inclusion of the ${\cal O} (a^4 )$ term in the shadow radius opens the door to more possibilities. 
An interesting exercise in this respect is to set some of the parameters to their Kerr values,  
$S_1 =1, \cM_2 = -1, c_{42} = c_{31}=0$. The resulting shadow radius now depends on the higher multipoles $\cS_3, \cM_4$,
\begin{align}
\frac{\bar{b}_{\rm ph}}{M} & \approx 5.19615 - 0.288675 a^2  +  \left (0.79105  \right.
\nonumber \\
& \left. \quad + 2.04045 \cM_4  + \,2.71703 \cS_3 \right )  a^4.
\label{baverM4S3}
\end{align}
The extent to which these two multipoles can decircularise the shadow can be understood by looking at the top panel of 
Fig.~\ref{fig:bHTcontour} where we show the contour plot of the fractional difference $ \Delta_b =  \bar{b}_{\rm ph}/b_{\rm Schw} -1$ (for $a=0.7$). 
For the $ (-1,3) \times (-3,1)$ `box' shown in the figure, the shadow radius can vary up to $\sim 40\%-50\%$ with respect to the circular 
Schwarzschild value. For the Kerr spacetime, $\cM_4^\rK =1, \cS_3^\rK = -1$, the deviation from a circular shadow is small, $| \Delta_b | \lesssim 0.1 $
(this is indicated in~Fig.~\ref{fig:bHTcontour} by a black dot). 
However, it is also clear that there is a high degree of degeneracy in the sense that  we can have $| \Delta_b | \ll 1 $ even for a markedly non-Kerr 
multipolar structure; an example is the point $(\cM_4, \cS_3) = (-1, 0.5)$. 

We can repeat the same exercise by varying the mass moments $\cM_2, \cM_4$ while setting the rest of the parameters 
equal to their Kerr values (i.e., $\cS_1= -\cS_3 =1, c_{31}=c_{42}=0$). The associated shadow radius is
\begin{align}
\frac{\bar{b}_{\rm ph}}{M} & \approx 5.19615 -  ( 1.06677 +  0.778098 \cM_2) a^2
\nn \\
&  + \left (\,  2.04045 \cM_4 -12.1183 - 21.335 \cM_2  \right.
\nn \\
& \left. - 11.1427 \cM_2^2 \, \right  ) a^4.
\label{baverM2M4}
\end{align}
Compared to~\eqref{baverM4S3}, this radius displays a markedly larger ${\cal O} (a^4)$ piece. This effectively limits how much we can 
vary $\cM_2$ while being consistent with the spin expansion character of $\bph$. For example, $\cM_2 \gtrsim 1$ could easily result in a 
negative radius, which is clearly unphysical. In order to account for this limitation, the corresponding contour plot of $\Delta_b$ is calculated 
for a somewhat lower spin, $a=0.5$, see middle panel of Fig.~\ref{fig:bHTcontour}.
The dominant role of $\cM_2$ as a shadow decircularisation factor is clearly visible in this figure: the contour lines are almost vertical and a variation 
across the $-2 \leq \cM_2 \leq 0.5$ range causes a reduction in the equatorial radius up to $\approx 40\%$ with respect to the Kerr radius. 
At the same time, however, we can tweak $\cM_4$ so that $\Delta_b$ remains close to its Kerr value even if $\cM_2$ deviates from Kerr. 
An example of this is the point $(\cM_2,\cM_4 ) = (-1.5,1.5)$ in Fig.~\ref{fig:bHTcontour}. 

Finally, we consider the shadow radius as a function of  $\cM_2, \cS_3$ with the rest of the parameters set equal to their Kerr values
($\cS_1^\rK= \cM_4^\rK =1, c_{31}=c_{42}=0$). The associated equatorial shadow radius is,
\begin{align}
\frac{\bar{b}_{\rm ph}}{M} & \approx 5.19615 - (1.06677 +  0.778098 \cM_2 ) a^2 + 
\nn \\
& + ( -2.68753 - 11.1427 \cM_2^2 + 7.39031 \cS_3 
\nn \\
& -16.6617  \cM_2  + 4.67328  \cM_2  \cS_3) a^4.
\end{align}
The quadrupole $\cM_2$ is still the dominant factor (leading to a similar deviation from a spherical shadow as the previous case) 
but its decircularising influence  is much more easily counteracted by a variation in $\cS_3$  than in $\cM_4$. 

The upshot of this analysis, and in combination with the results of the previous section, is that a quasi-circular shadow
does \emph{not} provide a reliable test of the no hair-theorem [Eq.~\eqref{nohair}] as different combinations of multipole moments 
beyond the quadrupole can produce the same, more or less, deviation from circularity as the Kerr spacetime. 
In terms of the dimensionless parameters $\cM_\ell, \cS_\ell$, this statement entails comparable magnitude shifts
away from the Kerr moments. 

\section{Shadow radius in a general stationary-axisymmetric spacetime}
\label{sec:axisym}

As a second case study of the shadow produced by a non-Kerr spacetime, in this section we consider a general
stationary-axisymmetric spacetime in GR with an arbitrary multipolar structure. This model is hinged on a double expansion; 
first with respect to the multipole moment order (which in practice is a slow rotation approximation if an $\ell$-pole is assumed to scale 
as $\sim J^\ell$) and subsequently with respect to $M/r$ for a given multipole moment order. 
The moments themselves are written with the Kerr and non-Kerr parts separated, 
\be
M_\ell = M_\ell^\rK + \delta M_{\ell},  \quad S_\ell = S_\ell^\rK + \delta S_\ell, \nn
\ee
and the former part is resummed so that the full Kerr part of the metric is recovered. This step ensures that the metric has the correct 
Schwarzschild and Kerr limits when, respectively,  rotation is turned off or the moments are set to their Kerr values.
In essence, this metric is what comes out when we superimpose Ryan's post-Newtonian metric of arbitrary multipole moments~\cite{Ryan:1995}
with the Kerr spacetime, making sure not to count terms twice (details on the construction of this spacetime will appear elsewhere). 
This makes it an ideal complement to the HT metric which is fully accurate at each spin order but is truncated to a lower multipole order than 
the metric discussed here. 
This spacetime is a direct descendant of the metric first introduced in Ref.~\cite{Maselli:2019qbf}, which was designed with an arbitrary set of 
multipole moments up to $M_4$ and the correct Schwarzschild limit. The present model improves on that earlier construction by pushing the 
expansion beyond $M_4$ and having the correct Kerr limit. Therefore, the new metric can be used as a parametrised non-Kerr solution within GR, 
with arbitrary deformations in the multipole moments.

A general stationary-axisymmetric vacuum spacetime can be constructed algorithmically via the Ernst potential formalism~\cite{ernst68}. 
For the case at hand, we employ the cylindrical-like Weyl-Papapetrou coordinates $\{t,\rho,z, \phi\}$ and the resulting line element takes the 
form~\cite{papapetrou53}
\begin{align}
ds^2  &=  -f \left(dt-\omega d\phi\right)^2  \nn
\\ 
& \quad + f^{-1} \left[ e^{2\gamma} (d\rho^2+dz^2)+\rho^2d\phi^2 \right].
\label{ds2axisym}
\end{align}
The three metric potentials $\{f, \omega, \gamma\}$ are functions of $\{ \rho, z \}$; they can be written in a separable form, comprising Kerr and 
non-Kerr parts (their full functional forms can be found in Appendix~\ref{sec:Ryanmetric}),
\begin{align}
 f &= f_\rK  +\delta f,
 \label{eq:axisym1}
 \\
 \omega &=  \omega_\rK + \delta\omega,
\label{eq:axisym2}
\\
e^{2  \gamma} &=  e^{2 \gamma_\rK} \left ( 1+ \delta \gamma \right ).
\label{eq:axisym3}
\end{align}
The functions $\{ f_\rK, \omega_\rK, \gamma_\rK \}$ represent the baseline Kerr solution in Weyl-Papapetrou coordinates and depend on $M$ 
and the dimensionless  Kerr spin parameter  $a=S_1/M^2$ (not to be confused with the Kerr parameter $a = J/M^2$ used elsewhere in the paper).  

The arbitrary deformations $\{ \delta M_\ell, \delta S_\ell \}$ in the multipole moments enter through the non-Kerr corrections 
$\{ \delta f, \delta \omega, \delta \gamma \}$. Once the metric is known, we can use the general formulae~\eqref{photonEq} and \eqref{bgeneral} 
to obtain $\rph$ and $\bph$ as expansions in the spin/multipolar order parameter $a$. The results take a somewhat simpler form when written 
in terms of the dimensionless multipolar deformation (this definition assumes that any deformation away from Kerr is rotation induced),
\be
\delta \cM_\ell \equiv \frac{\delta M_\ell}{a^\ell M^{\ell+1}}, \qquad  \delta \cS_\ell \equiv \frac{\delta S_\ell}{a^\ell M^{\ell+1}}.
\ee
After a numerical evaluation, we arrive at the following expressions:
\begin{widetext}
\begin{align}
\frac{\bph}{M} &=3 \sqrt{3}-2 a  -1.19772 \Big ( 0.646842 \, \delta \cM_2 +0.24102  \Big )  a^2 
-0.0800267  \Big( 16.1044  \, \delta \cM_2  -4.47113 \, \delta \cS_3 
\nn \\
& + \, 1.85123 \Big )  a^3 
+ 6.16711  \Big ( 0.0321565 \, \delta \cM_4 -0.0838228\, ( \delta \cM_2) ^2 -0.342846 \, \delta \cM_2   
+ 0.126939\, \delta \cS_3
\nn\\ 
& -\, 0.0151695 \Big ) a^4  -31.0214   \Big( 0.0951536\, ( \delta \cM_2)^2  -0.0199417 \, \delta \cM_2\, \delta \cS_3 +0.163008\, \delta \cM_2  
\nn \\
& - \, 0.0189\, \delta \cM_4 -0.0357977 \, \delta \cS_3 + 0.00330828\, \delta \cS_5 + 0.00212252 \Big ) a^5  +O\left(a^6\right),
 \\
 \nn \\
\frac{\bar{b}_{\rm ph}}{M} &=3 \sqrt{3}  -1.19772 \Big ( 0.646842 \,\delta \cM_2 + 0.24102  \Big ) a^2  
 + 6.16711  \Big (-0.0838228\, ( \delta \cM_2 )^2 -0.342846\, \delta \cM_2 
 \nn \\
&  \quad +\, 0.0321565 \,\delta \cM_4 + 0.126939 \, \delta \cS_3 - 0.0151695 \Big ) a^4  +O\left(a^6 \right).
\end{align}
\end{widetext} 
As pointed out earlier, $\bar{b}_{\rm ph}$ is given by the same expression as $\bph$ after removing the odd powers of $a$.

These results look very similar to those obtained in the HT spacetime [e.g., Eq.~\eqref{bphMM2}] but, as expected, are not identical. 
By construction, the general metric of this section is a function of the multipole moments only, while the HT metric contains the structure constants $C_{nm}$.
As we have seen, it is not possible to replace all of these parameters with $M_\ell, S_\ell$. 
The proximity of the two $\bar{b}_{\rm ph}$ results can be gauged if we calculate the $\cM_2$ that makes the shadow circular at ${\cal O} (a^2)$. 
We find 
\be
\delta \cM_2 \approx - 0.373 ~\Rightarrow ~ \cM_2 \approx -1.373 \cS_1^2,
\ee
which lies very close to our earlier HT result~\eqref{M2circHT}.

\begin{figure}[htb!]
\begin{center}
\includegraphics[width=0.36\textwidth]{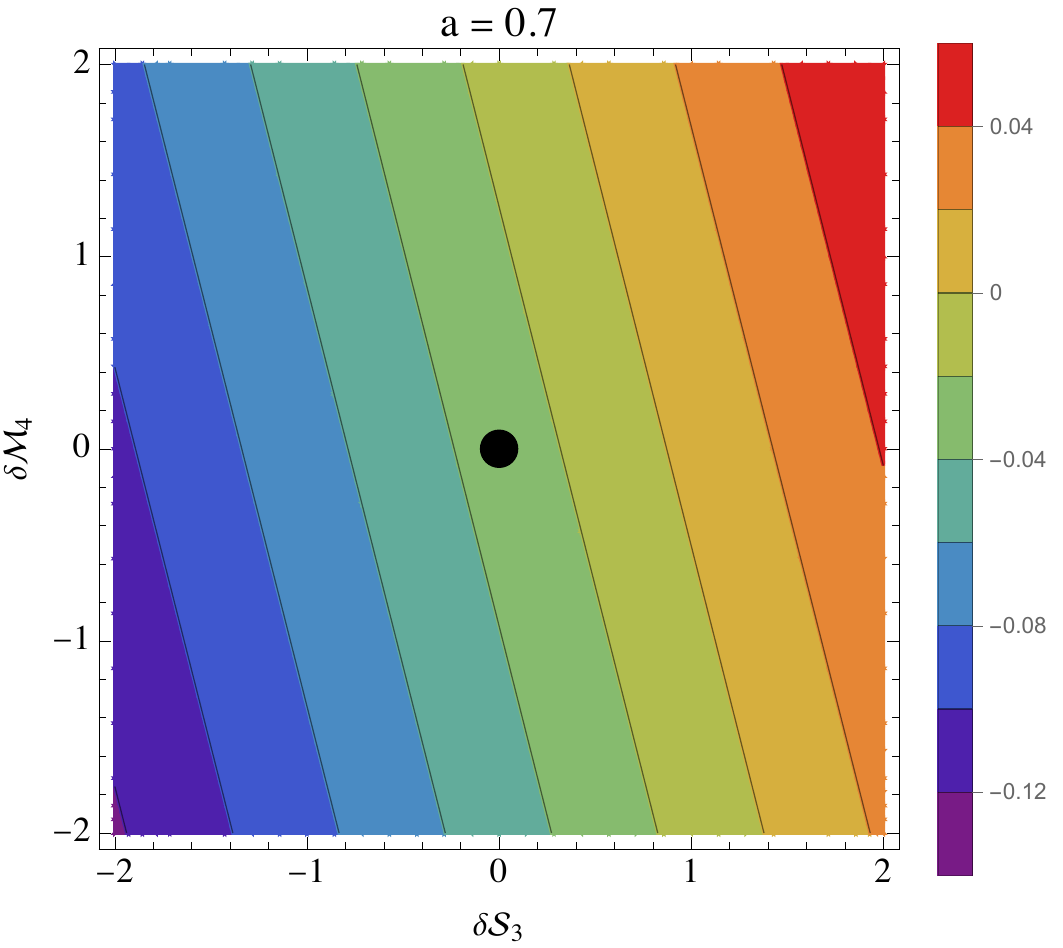}
\\ \vspace{0.45cm}
\includegraphics[width=0.36\textwidth]{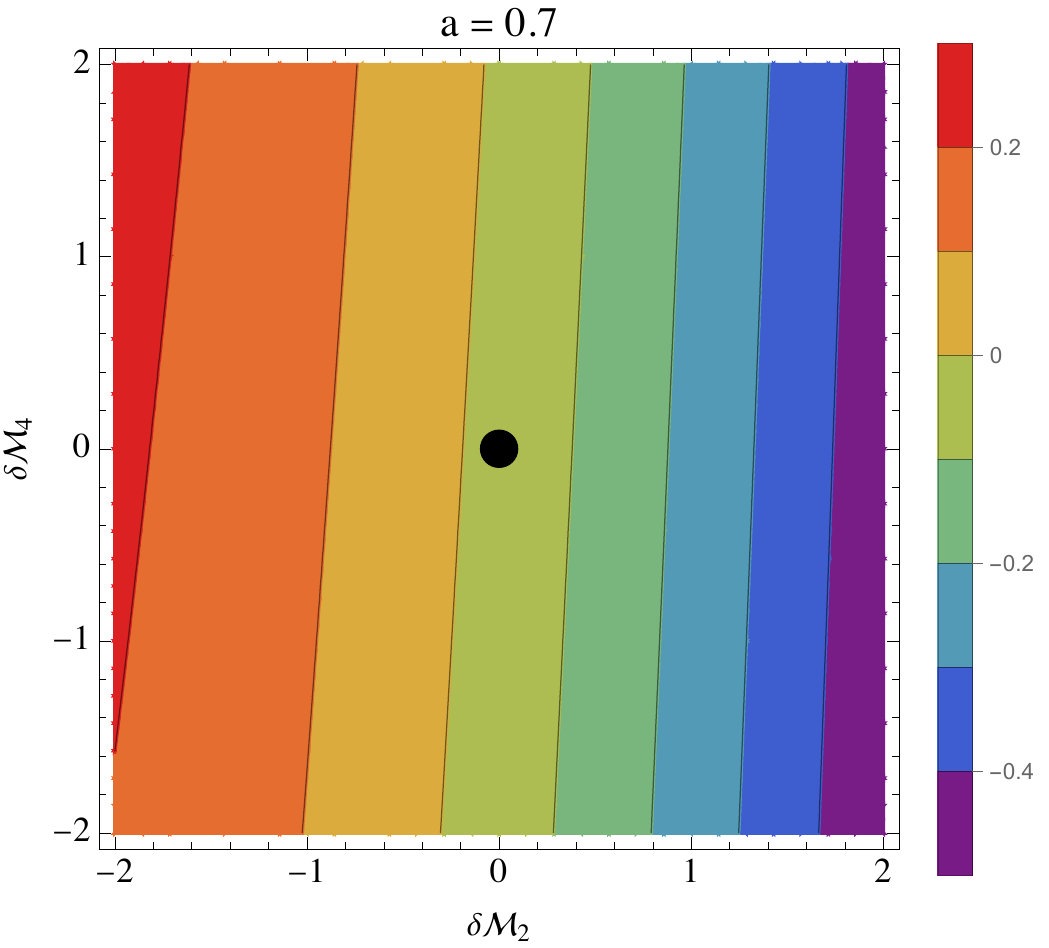}
\\ \vspace{0.45cm}
\includegraphics[width=0.36\textwidth]{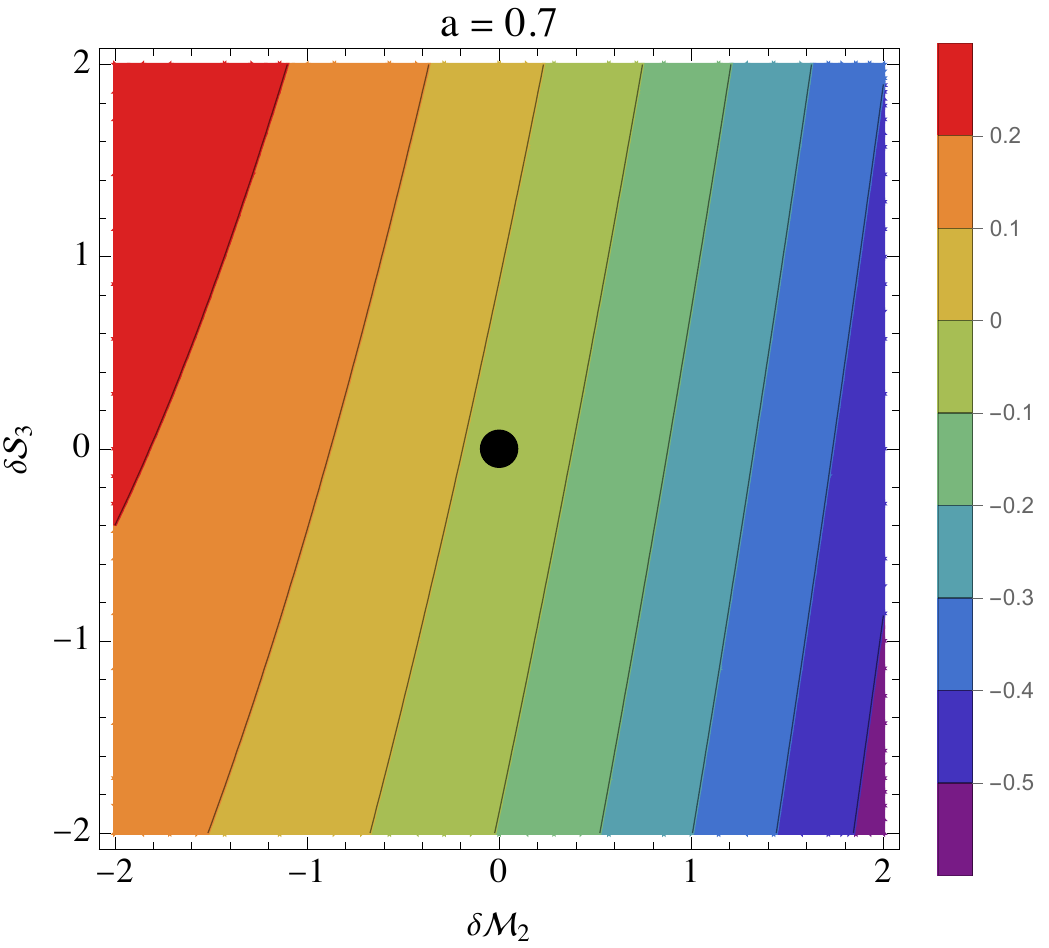}
\end{center}
\caption{Contour plot of the equatorial shadow radius expressed as a fractional difference $ \Delta_b =  \bar{b}_{\rm ph}/b_{\rm Schw} -1 $
with respect to the circular Schwarzschild radius. Top panel:  $\Delta_b$ as a function of the $\delta \cS_3, \delta \cM_4$ deviation 
parameters, assuming $\delta \cM_2= 0$ (i.e. $\cM_2 = \cM_2^\rK = -1$). 
Middle panel: $\Delta_b$ as a function of the $\delta \cM_2, \delta \cM_4$ deviation parameters, assuming $\delta \cS_3 = 0$ 
(i.e., $\cS_3 = \cS_3^\rK =-1$). 
Bottom panel: $\Delta_b$ as a function of the $\delta \cM_2, \delta \cS_3$ deviation parameters, assuming $\delta \cM_4 = 0$ 
(i.e., $\cM_4 = \cM_4^\rK =1$). 
In all cases, the spin is set to $a=0.7$. The black dot marks the Kerr limit of $\Delta_b$.}
\label{fig:shadowG_frac}
\end{figure}

Without further ado we calculate the degree of deviation of this new $\bar{b}_{\rm ph}$ from the circular radius $b_{\rm Schw}$ and 
the Kerr shadow radius. In Fig.~\eqref{fig:shadowG_frac} we show contour plots of the fractional difference  $ \Delta_b =  \bar{b}_{\rm ph}/b_{\rm Schw} -1$ 
in the shadow radius for a dimensionless spin $a=0.7$ and three complementary choices for the multipole moments: 
(i) We fix $M_2 = M_2^\rK$ and vary $\delta \cS_3,  \delta \cM_4$ (top panel). In both cases, the Kerr limit 
is indicated by a black dot and is clearly seen to represent a nearly circular shadow for the chosen spin value.
(ii) We fix $S_3 = S_3^\rK$ and vary $\delta \cM_2, \delta \cM_4$ (middle panel).
(iii) We fix $M_4 = M_4^\rK$ and vary $\delta \cM_2, \delta \cS_3$ (bottom panel).

Considering first $\bar{b}_{\rm ph}$ as a function of the higher multipoles $M_4, S_3$, we can observe a much smaller
variation in $\Delta_b$ ($\lesssim 15\% $) with respect to the previous HT case, for the same$(-2,2)\times (-2,2)$ `box' and spin. 
In other words, for the spacetime discussed in this section the shadow remains nearly circular for a wide range of deviation
around the $S_3, M_4$ moments. The degree of decircularisation of the shadow becomes much higher 
($\lesssim 50 \%$ for the range shown in Fig.~\ref{fig:shadowG_frac}), and similar to the one found in the HT spacetime, 
when the mass quadrupole $M_2$ is varied. In all cases, however, we can draw the same conclusion as before: we can shift two 
or more multipoles away from Kerr while maintaining a Kerr-like shadow radius.  
%

\section{The Johannsen metric: multipole moments and shadow radius}
\label{sec:Jmetric}

\subsection{Extracting the multipole moments}
\label{sec:Jmoments}

The widely used Johannsen metric~\cite{Johannsen2013} (hereafter `J metric', see Appendix~\ref{sec:Jexpansions}) is an example of a deformed Kerr spacetime 
with the characteristic property of separability, that is, it admits a Carter-like constant. The purpose of this section is to study the multipolar structure of the J spacetime 
in relation to its black hole shadow.

In reality, this is an ill-defined objective; the formal calculation of a spacetime's multipole moments in a given theory of gravity requires the use of 
field equations and appropriate asymptotic conditions~\cite{Pappas:2012, pappas15a}. No such field equations are available for the J spacetime for the simple reason that
it is not a solution of GR or any other known theory of gravity. 

In the absence of field equations that could determine the multipole moments, we follow a more practical approach by 
deriving post-Newtonian expansions for the orbital frequencies of a test body in a small eccentricity/inclination orbit.
These expressions can be compared against the ones obtained by Ryan~\cite{Ryan:1995} for a stationary-axisymmetric spacetime 
of arbitrary multipolar structure in GR with the aim of extracting some information about the multipolar structure of the J spacetime 
from the  coefficients of the post-Newtonian expansion.
Two of the body's equations of motion are
\be
u^t = \frac{1}{\cD} \left (  g_{t\varphi} L + g_{\varphi\varphi} E \right ), ~
u^\varphi = -\frac{1}{\cD} \left (  g_{t\varphi} E  + g_{tt} L  \right ).
\ee
In addition, we can define an effective potential (not to be confused with the one used earlier for null geodesics)
\begin{align}
 & g_{rr} (u^r )^2  + g_{\theta\theta} \left ( u^\theta \right )^2 
  = \frac{1}{\cD} \left (\, g_{tt} L^2 + 2 g_{t\varphi} L E  + g_{\varphi\varphi}  E^2\, \right )
 \nonumber  \\
 & -1  \equiv   V_{\rm eff} (r,\theta,E,L).
\label{norm_gen}
\end{align}
Circular equatorial orbits ($\theta=\pi/2$) are required to solve
\be
V_{\rm eff} (r_0, E, L) = V_{\rm eff}^\p (r_0, E, L) = 0,
\ee
where $r_0$ denotes the orbital radius. These two conditions can be solved for any pair of parameters; in
the present case, we are interested in the angular frequency $\Omega = u^\varphi/u^t$. Making the substitutions 
\be
E= - u^t \left  ( g_{tt} + \Omega g_{t\varphi} \right ), \qquad  L=  u^t \left  ( g_{t\varphi} + \Omega g_{\varphi\varphi} \right ).
\ee
we can solve the above system for $\{ \Omega, u^t \}$. The $V^\p_{\rm eff} =0$ equation becomes a binomial,
\be
g_{\varphi\varphi}^\p \Omega^2 + 2 g_{t\varphi}^\p \Omega + g_{tt}^\p = 0, 
\ee
with solutions
\be
\Omega_{\pm} = \frac{1}{g_{\varphi\varphi}^\p} \left ( -  g_{t\varphi}^\p \pm \sqrt{  g_{t\varphi}^{\p 2} -  g_{tt}^\p  g_{\varphi\varphi}^\p }\right ).
\ee
The condition $ V_{\rm eff} = 0$ yields the `redshift' formula,
\be
u^t = \left (- g_{tt}  -g_{\varphi\varphi} \Omega^2 - 2 g_{t\varphi} \Omega  \right )^{-1/2}.
\ee
A perturbed circular orbit is characterised by a small inclination/eccentricity. In addition to $\Omega$, we now have to consider
the epicyclic orbital frequencies $\Omega_r, \Omega_\theta$. To leading order in the perturbation, these are given by,
\be
\Omega_r = \frac{1}{u^t} \sqrt{-\frac{V_{\rm eff}^{\pp} }{2 g_{rr}}}, \qquad  
\Omega_\theta = \frac{1}{u^t} \sqrt{-\frac{ \partial_\theta^2 V_{\rm eff}}{2 g_{\theta\theta}}},
\label{epicyclicOm}
\ee
where the right-hand-side terms are to be evaluated for the unperturbed circular equatorial orbit.  
The orbital precession frequencies are defined as
\be
\omega_i \equiv \Omega -\Omega_i, \qquad i = \{r,\theta\}.
\label{precessOm}
\ee
These general formulae can now be applied to the J metric (the detailed form of this metric is given in Appendix~\ref{sec:Jexpansions}). 
Using $\epsilon$ as a collective bookkeeping parameter for the metric's deformation parameters $ \{ \alpha_{13}, \alpha_{22}, \alpha_{52}, \varepsilon_3 \}$, 
we can write $\Omega$ as an expansion in $\epsilon$ and $M/r_0$. The result is Eq.~\eqref{Omexpan} in Appendix~\ref{sec:Jexpansions}.
The $\Omega$ expansion can be inverted and furnishes $M/r_0$ as an expansion in $\epsilon$ and the orbital velocity $v^3 = M \Omega$, see Eq.~\eqref{vexpan}.

This latter result can be subsequently used in Eqs.~\eqref{epicyclicOm}-\eqref{precessOm} to obtain expansions for the precession frequencies.
For the normalised frequencies  $\tilde{\omega}_i = \omega_i /\Omega$ we find (here $a = J/M^2$ is the Kerr spin parameter)
\begin{widetext}
\begin{align}
\tilde{\omega}_r & = 3 v^2 - 4a  v^3  + \frac{1}{2}  \left [ 3 \left ( 3 + a^2 \right ) + \epsilon ( 6\alpha_{13} -\alpha_{52} - 3\varepsilon_3 ) \right ] v^4
+ \frac{1}{2}  \left [ 27 + 17 a^2 + 3 \epsilon ( \alpha_{52} -4\alpha_{13} + 5 \varepsilon_3 ) \right ] v^6 
\nonumber \\
&  -5a  \left ( 2 + \epsilon \alpha_{22} \right ) v^5  + a \left [ -4 \left ( 12 + a^2 \right ) + \frac{1}{3} \epsilon ( 3 \alpha_{13} -21\alpha_{22} \
- 8\alpha_{52} -15 \varepsilon_3 )  \right ] v^7  
+ {\cal O} \left (\epsilon^3, v^8 \right ), 
\\
\nonumber \\
\tilde{\omega}_\theta & =  2a v^3 - \frac{3}{2} a^2 v^4 + \epsilon  a \alpha_{22}  v^5
+ 4a^2 v^6 + a \left [ - 5a^2 + \epsilon ( 3\varepsilon_3 -5 \alpha_{13} ) \right ] v^7 
+ {\cal O} \left (\epsilon^3, v^8 \right ).
\end{align}
\end{widetext}
Note that no ${\cal O} (\epsilon^2)$ terms appear in these results. These expressions can be compared against 
the multipole moment expansions of Ref.~\cite{Ryan:1995} for $\tilde{\omega}_i$, derived for orbits in an 
arbitrary axisymmetric-stationary spacetime in GR:
\begin{widetext}
\begin{align}
( \tilde{\omega}_r)_{\rm GR}  & = 3v^2 - \frac{4 S_1}{M^2} v^3 + \frac{3}{2} \left ( 3-\frac{M_2}{M^3} \right ) v^4
- 10 \frac{S_1}{M^2} v^5 + \left ( \frac{27}{2} - 2  \frac{S_1^2}{M^4} - \frac{21}{2}  \frac{M_2}{M^3}  \right ) v^6 
\nonumber \\
& \quad + \left ( -48  \frac{S_1}{M^2}  - 5   \frac{S_1}{M^2} \frac{M_2}{M^3} + 9 \frac{S_3}{M^4}   \right ) v^7 
 + {\cal O} \left ( v^8 \right ), 
\\
\nonumber \\
( \tilde{\omega}_\theta )_{\rm GR} & = \frac{2 S_1}{M^2} v^3 + \frac{3}{2} \frac{M_2}{M^3} v^4 + 
\left ( 7 \frac{S_1^2}{M^4} + 3 \frac{M_2}{M^3} \right ) v^6 
+ \left (  11 \frac{S_1}{M^2} \frac{M_2}{M^3} - 6 \frac{S_3}{M^4} \right ) v^7 + {\cal O} \left ( v^8 \right ).
\end{align}
\end{widetext}
Inspection of the first four $v^n$ pairs leads to the following identifications [the first (second) entry in each line corresponds to 
 $\tilde{\omega}_r$ ($\tilde{\omega}_\theta$)]:
\begin{align}
{\cal O} (v^3): & ~ S_1 = a M^2,
\\
{\cal O} (v^4): &~ \frac{M_2}{M^3} =  a^2  -\frac{1}{3} \epsilon( 6 \alpha_{13} - \alpha_{52} - 3 \varepsilon_3 ), 
\nonumber \\
& ~ \frac{M_2}{M^3} = - a^2
\\
{\cal O} (v^5): & ~ \mbox{unbalanced terms}~ \left \{ 5 a \alpha_{22}, -a  \alpha_{22} \right \},
\\
{\cal O} (v^6): &~ \frac{M_2}{M^3} = - a^2 - \frac{1}{7}  \epsilon ( \alpha_{52} -4 \alpha_{13} + 5 \varepsilon_3 ),
\nonumber \\
& ~  \frac{M_2}{M^3} = - a^2.
\end{align}
The predicted quadrupole moment from the ${\cal O} (v^4)$ terms is unique and identical to the Kerr value, $M_2 = -a^2 M^3$,  
provided we set $\alpha_{52} = 6 \alpha_{13}  - 3 \varepsilon_3$.
Similarly, in order to avoid inconsistency in the ${\cal O} (v^5) $ terms we must set $\alpha_{22} = 0$. Using this information in the ${\cal O} (v^6)  ~\mbox{terms}$
fixes one more parameter, $\alpha_{13} = - \varepsilon_3$. Finally, the highest-order terms lead to
\be
{\cal O} (v^7): ~ \frac{S_3}{M^4} = - a^3  +  2a  \varepsilon_3,  \quad \frac{S_3}{M^4} = - a^3  - \frac{4}{3} a \varepsilon_3.
\ee
We have thus hit an inconsistency wall: setting $\varepsilon_3 =0$ makes the rest of the parameters vanish and the metric
reduces to Kerr. Based on our earlier comment this result is hardly surprising; the J metric is not a solution of the vacuum GR equations
and therefore we should not have expected a fully consistent correspondence with  Ryan's post-Newtonian multipolar expansion.


\subsection{Shadow radius in the Johannsen metric}
\label{sec:Jshadow}

The second (and last) part of our discussion of the J spacetime is concerned with the black hole's shadow (previous work 
on the subject can be found in~\cite{Johannsen2013ApJ}). 

The light ring radius and impact parameter are calculated with the help of Eqs.~\eqref{photonEq} and \eqref{bgeneral}, 
without expanding in the deformation parameters. As it turns out both quantities depend only on two parameters, 
$\{\alpha_{13}, \alpha_{22}\}$. 

\begin{figure}[htb!]
\begin{center}
\includegraphics[width=\columnwidth]{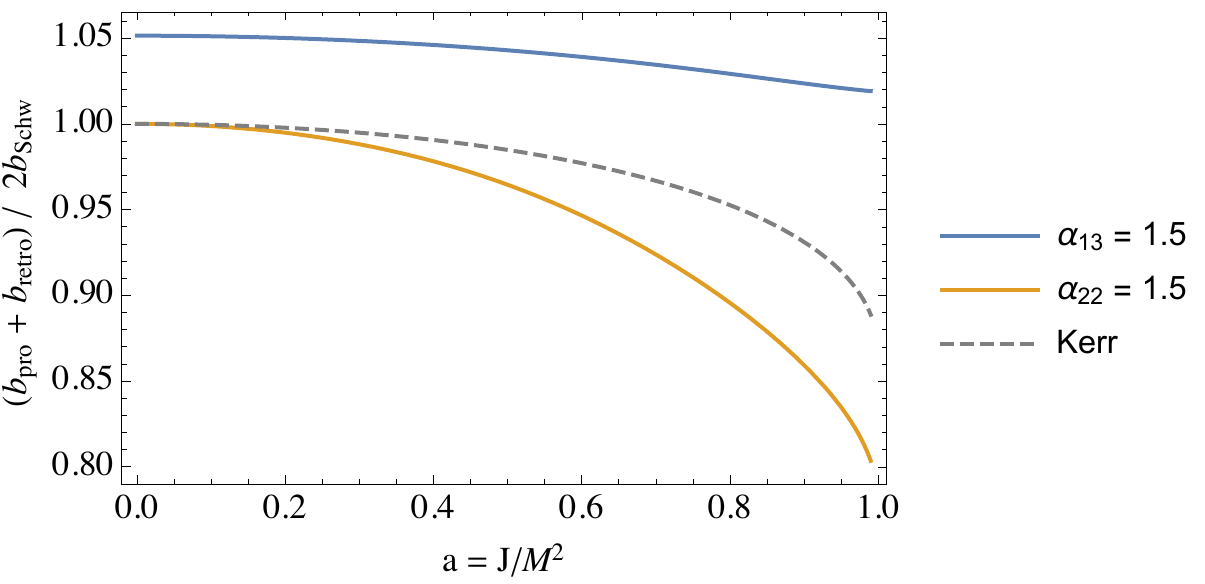}
\\ \vspace{0.5cm}
\includegraphics[width=\columnwidth]{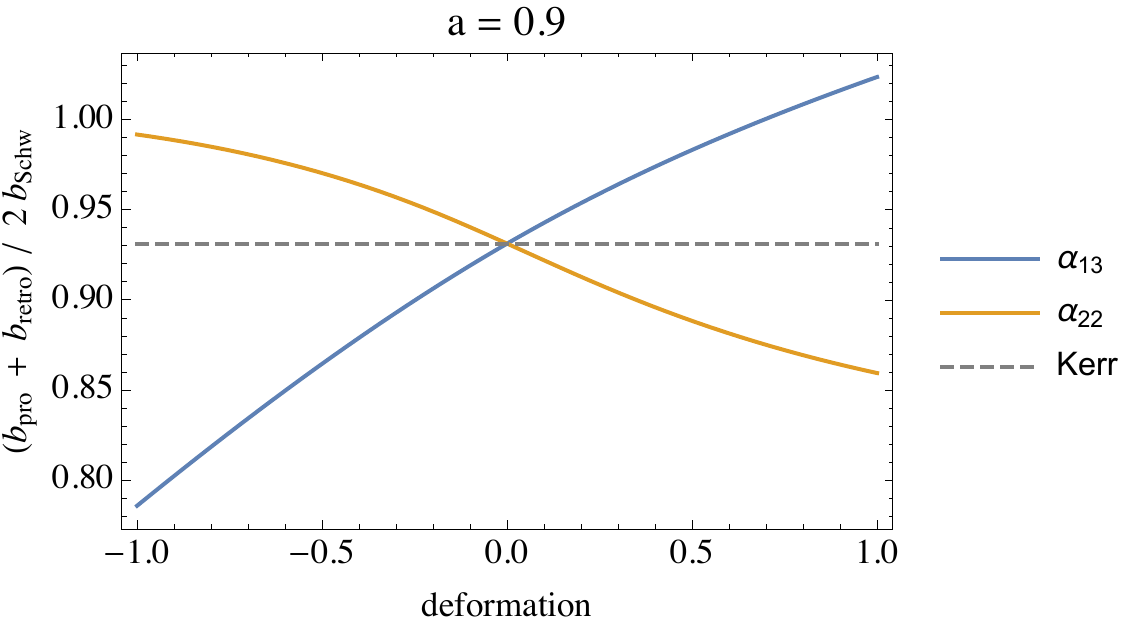}
\end{center}
\caption{The normalised equatorial shadow radius $\bar{b}_{\rm ph}/b_{\rm Schw}$ of a black hole 
in the J spacetime (solid curves). Top panel: as a function of the dimensionless spin $a=J/M^2$ for fixed deformation parameters 
$\alpha_{13} = \alpha_{22} = 1.5$.
Bottom panel: as a function of  $\alpha_{13},  \alpha_{22}$ for $a=0.9$. For comparison, we also show the Kerr equatorial shadow 
(dashed curves). }
\label{fig:Jshadow}
\end{figure}

As already discussed, the equatorial shadow radius is given by the prograde-retrograde averaged impact parameter,
$ \bar{b}_{\rm ph} =  \left ( b_{\rm pro} + b_{\rm retro} \right )/2$. The ratio $\bar{b}_{\rm ph}/b_{\rm Schw}$ is shown in 
Fig.~\ref{fig:Jshadow} as a function of the dimensionless spin $a$ (top panel) and of the deformation parameters (bottom panel). 

These results suggest that not all deformation parameters lead to the same degree of deviation from a circular-shaped 
shadow. In the example shown in Fig.~\ref{fig:Jshadow},  the departure from the $a=0$ shadow radius is negligible when
$\alpha_{13} > 0$; in contrast, a positive $\alpha_{22}$ makes the shadow markedly less circular than its Kerr counterpart 
for any spin $a \gtrsim 0.5$.


\section{Concluding remarks}
\label{sec:conclusions}

The purpose of this paper was to build a bridge between the no-hair theorem's multipolar relation~\eqref{nohair} and the equatorial radius of the 
shadow cast by a Kerr black hole mimicker (this term includes ultracompact objects with a material surface but negligible surface emission as well as 
non-Kerr black holes) within the framework of GR.  

With the help of two stationary-axisymmetric vacuum metrics endowed with rotation and an arbitrary set of multipole moments, we have
been able to produce analytic formulae, in the form of slow rotation expansions, for the shadow radius as an explicit function 
of the moments themselves (or, equivalently, their deviation from the Kerr moments). These formulae have subsequently allowed us to explore the 
variation in the equatorial shadow radius (which serves as a measure of the shadow's degree of circularity or lack thereof) in the multipole moment 
parameter space. 
Our model represents a significant improvement on previous work on the subject~\cite{Johannsen2010ApJ, Broderick2014ApJ} by 
making use of GR spacetimes with rotation several non-Kerr moments beyond quadrupole order and by directly expressing the shadow radius
as a function of these multipoles.

The main result of our analysis is rather clear: a quasi-circular shadow which typically characterises a Kerr black hole could also be the result of 
light propagating in a spacetime with a significant degree of  deviation from the Kerr multipolar structure.  Therefore, addressing the question posed in the title
of this paper, we can say that a black hole shadow \emph{ may not} necessarily be a reliable test of the no-hair theorem. 
This conclusion appears to be at odds with Refs.~\cite{Johannsen2010ApJ, Broderick2014ApJ}, but there is no
real contradiction here because those earlier papers assumed a spacetime with a single non-Kerr multipole moment (the quadrupole). 
This would correspond to a variation along a horizontal line passing through the `Kerr point' in some of the panels of Figs.~\ref{fig:bHTcontour} and
\ref{fig:shadowG_frac}. 
It should be emphasised that our analysis does not imply that a black hole shadow image cannot be used as
a test of the Kerr spacetime. In fact, the results shown in aforementioned figures do allow for an appreciable deviation from a 
Kerr-like shadow if one `moves' along a suitable direction in the multipole moment plane.  The real impact of our results is to
weaken the link between the moments and the shadow shape in the sense that an observation of a Kerr-like shadow 
does not necessarily imply a small deviation from the Kerr moments.

A secondary implication of our results is that a nearly circular shadow does \emph{not} appear to be an exclusive property of separable stationary-axisymmetric 
metrics with a Carter-like geodesic constant of motion. Indeed, the two spacetimes explored in this paper (and in contrast to the Kerr and J metrics) 
are not separable with respect to the Hamilton-Jacobi equation for point particle motion.   

It is conceivable that the analysis presented in this paper could be extended to non-GR theories of gravity, thus enabling 
a connection between the black hole shadow shape and the multipole moments of genuine non-Kerr black holes. 
For example, Refs.~\cite{pappas15a,pappas15b} have developed an Ernst formalism-based framework for the definition and calculation of 
multipole moments of asymptotically flat, stationary-axisymmetric spacetimes in scalar-tensor gravity. The resulting metric could be
used in the same way as the two metrics of the present paper for the calculation of a black hole shadow radius in a slow rotation
approximation. It is unclear, however, if the same method could work equally well for other theories of gravity whose field equations cannot
be reduced to an Ernst potential formalism.


\acknowledgments

K. G. acknowledges support from research Grant No. PID2020-1149GB-I00 of the Spanish Ministerio de Ciencia e Innovaci{\'o}n.  

\appendix

\section{A general stationary and axisymmetric metric with a Kerr limit}
\label{sec:Ryanmetric}

This appendix provides the full functional form of the metric potentials $\{f, \omega, \gamma \}$ introduced in Section~\ref{sec:axisym}, 
see Eqs.~\eqref{ds2axisym}-\eqref{eq:axisym3}. The Kerr part of the metric, expressed in Weyl-Papapetrou coordinates, is
\begin{align}
                      f_{\rm K}(\rho,z)&= 1-\frac{2 \left(1+x \sqrt{1-a^2}\right)}{a^2 y^2+\left(1+x\sqrt{1-a^2}\right)^2},
                      \\
                      \nn \\
                      \omega_{\rm K}(\rho,z) &= \frac{2a(y^2-1)\left(1+x \sqrt{1-a^2}\right)}{(x^2-1)+a^2(y^2-x^2)},
                      \\
                      \nn \\
                      \gamma_{\rm K}(\rho,z) &=\frac{1}{2}\log \left[ \frac{(x^2-1)+a^2(y^2-x^2)}{\left(1-a^2\right)\left(x^2-y^2\right)}\right],
\end{align}
where $a=S_1/M^2$ is the dimensionless Kerr spin parameter and
\begin{align}
                        x&=\frac{r_{+} + r_{-}}{2M\sqrt{1-a^2}}, \qquad
                        y =\frac{r_{+} - r_{-} }{2M\sqrt{1-a^2}},
                        \\
                        \nn \\
                        r_{\pm} &= \sqrt{\left(M\sqrt{1-a^2} \pm z\right)^2+\rho^2}.
\end{align}
The non-Kerr portion of the metric consists of the functions $\{ \delta f, \delta \omega, \delta \gamma \}$. (While $\delta \gamma$
does not enter in any of the analytic calculations for the light ring and the impact parameter, we show it here for completeness.)
They depend on the deviations $ \{\delta M_\ell, \delta S_\ell \}$ off the Kerr multipole moments as well as
their dimensionless combinations,
\be
\lambda_M \equiv \frac{ ( \delta M_2)^2}{M \delta M_4}, \qquad \lambda_S \equiv \frac{\delta S_3\delta M_2}{M \delta S_5}.
\ee
The three functions are given by the significantly lengthier expressions
\begin{widetext}
\begin{align}
             \delta f&=   \left[1-\frac{2   M}{  \rho}+\frac{16 M^2-63 z^2}{14  \rho^2} 
                        +\frac{5   M \left(M^2+14 z^2\right)}{7  \rho^3} -\frac{M^4+\frac{2}{7} M^2 z^2-\frac{75 z^4}{8}}{  \rho^4}
                          - \frac{  M \left(484 M^2 z^2+9 M^4+672 z^4\right)}{28   \rho^5}  \right 
                          .\nn\\
     &+\frac{  M \left(1250 M^4 z^2+4920 M^2 z^4+11 M^6+2464z^6\right)}{56   \rho^{7}}
   -\frac{  M \left(12376 M^6 z^2+88352 M^4 z^4+122560 M^2 z^6-3307 M^8+31360 z^8\right)}{448 \rho^{9}}\nn\\
     &\left.-\frac{  \left(13288 M^6 z^2+6736 M^4 z^4-51744 M^2 z^6+3984 M^8-19845 z^8\right)}{896 \rho^{8}} 
      + \frac{  \left(856 M^4 z^2-1568 M^2 z^4+90 M^6-1715 z^6\right)}{112 \rho^6} \right] \frac{\delta M_2 \varepsilon ^2}{\rho^3} \nn\\   
             &-\left[\frac{3}{4} -\frac{\left(\lambda_M +3\right) M}{2 \rho} +\frac{\left(7+3 \lambda_M \right) M^2 - 75  z^2}{8 \rho^2} 
            -\frac{\left(544 \lambda_M +315\right) M^4+\left(2212-2604 \lambda_M \right) M^2 z^2-17150 z^4}{448   \rho^4} \right.\nn\\
  & -\frac{\left(\frac{487}{392} \lambda_M +\frac{9}{32}\right)M^5+\left(\frac{1003}{28} \lambda_M +\frac{43}{2}\right) M^3 z^2+\left(\frac{173}{2} 
  +\frac{39}{2}\lambda_M \right)M z^4}{\rho^{5}}+\frac{\left(31 \lambda_M +14 \right)M^3+\left(546+126 \lambda_M \right)M z^2}{28 \rho^3} \nn\\
   &+\frac{\left(55576 \lambda_M +18914\right) M^6+\left(40768 \lambda_M +70413\right) M^4 z^2
   -\left(41062+373086 \lambda_M \right) M^2 z^4-648270  z^6}{6272   \rho^{6}}
   \nn\\
   &\left.-\frac{\left(\frac{2843}{224}\lambda_M +\frac{297}{64}\right)M^7-\left(\frac{33287}{392} \lambda_M 
   +\frac{833}{32} \right)M^5 z^2 -\left(\frac{547}{2} \lambda_M +180 \right) M^3 z^4-\left(\frac{505}{2}
   +\frac{115}{2} \lambda_M \right)M z^6}{  \rho^{7}} \right]   \frac{\delta M_4 \varepsilon ^4}{\rho^5} +\mathcal{O}{\left(\varepsilon^6\right)}, 
\end{align}
%
\begin{align}
 \delta \omega&= 
   \left[1+\frac{3}{2 \rho} +\frac{ \left(M^2-15 z^2\right)}{2 \rho^2}-\frac{2 M \left(M^2+18 z^2\right)}{3 \rho^3}
   +\frac{ \left(78 M^2 z^2-14 M^4+525 z^4\right)}{24 \rho^4} +\frac{\frac{19  M^5}{96}+18 M^3 z^2+39 M z^4}{\rho^5} \right.
    \nn\\
   &+\frac{ \left(3780 M^4 z^2-30838 M^2 z^4+315 M^6-39690 z^6\right)}{864 \rho^6}-\frac{ M \left(1568 M^4 z^2+10816 M^2 z^4+9 M^6+8640 z^6\right)}{96 \rho^{7}}
   \nn\\
   &\left.+\frac{  M^2 z^2 \left(1634 M^2 z^2+226 M^4+1551 z^4\right)}{33 \rho^{8}}\right] \frac{\delta S_3 \varepsilon ^3}{\rho^3}  
   -  \left[\frac{3}{4} +\frac{(2  \lambda_S +5) M}{4 \rho} +\frac{(5+2  \lambda_S )M^2-105 z^2}{8 \rho^2}\right.
   \nn\\
    &-\frac{\frac{15}{14}  \lambda_S M^3+\frac{5 M^3}{16}+\frac{45}{2} M z^2+3 \lambda_S M z^2}{\rho^3} 
   - \frac{\frac{73}{84} \lambda_S M^4+\frac{25  M^4}{64} +\frac{75}{16} M^2 z^2-\frac{129}{8} \lambda_S M^2 z^2-\frac{19525 z^4}{288}}{\rho^4}
   \nn\\
   & -\frac{-\frac{95}{84}  \lambda_S M^5-\frac{9 M^5}{64}-41  \lambda_S M^3 z^2-\frac{35}{2} M^3 z^2-\frac{497}{4} M z^4-\frac{25}{2}  \lambda_S M z^4}{\rho^{5}} 
   \nn\\
   &\left.- \frac{M z^2 \left(2670  \lambda_S M^3+385 M^3+2135  M z^2+7532  \lambda_S M z^2\right)}{77 \rho^{6}} \right] 
      \frac{\delta S_5\varepsilon ^5}{\rho^5}+\mathcal{O}\left(\varepsilon^7\right),
\end{align}
%
\begin{align}
         \delta \gamma&= \left [ \frac{3}{2}-\frac{3\left(9 M^2+56
   z^2\right)}{14 \rho^2}+\frac{\frac{463  M^4}{224}+\frac{243}{7}M^2 z^2+39z^4}{\rho^4} -\frac{7408 M^4 z^2+22040 M^2 z^4+239 M^6+10080 z^6}{112 \rho^6}\right ]
    \frac{ M \delta M_2\varepsilon ^2}{\rho^4}\nn\\
   &-\left[\frac{3 \lambda_M +5}{4} + \frac{(857 \lambda_M +350)M^2+(5040+3024 \lambda_M)z^2}{224 \rho^2}
   -\frac{\lambda_M M^2  \left(7408 M^4 z^2+22040 M^2 z^4+239 M^6+10080 z^6\right)^2}{25088 \rho^{14}}
   \right.\nn\\
   &
   +\frac{ \left (\frac{6271 }{784}   \lambda_M+\frac{109}{64} \right )M^4+ \left (\frac{731}{7}  \lambda_M +50 \right ) M^2 z^2
   + \left (\frac{505}{4}  +\frac{315}{4}  \lambda_M \right )z^4}{\rho^{4}} +\frac{3 \lambda_M M^2 \left(78624 M^2 z^2+5185 M^4+136416 z^4\right)}{3136 \rho^{6}}
   \nn\\
   &-\frac{\lambda_M M^2 \left(22571328 M^6 z^2+186291648 M^4 z^4+390254592 M^2 z^6+627361 M^8+184697856 z^8\right)}{100352 \rho^{10}}
   \nn\\
   &+\frac{\lambda_M M^2 \left(5288368 M^8 z^2+69897032 M^6  z^4+240766368 M^4 z^6+270923520 M^2 z^8+110657 M^{10}+88058880 z^{10}\right)}{25088 \rho^{12}}
   \nn\\
   &\left. +\frac{3 \lambda_M M^2 \left(199624 M^4 z^2+822640 M^2 z^4+7513 M^6+630336 z^6\right)}{3136 \rho^{8}}\right] \frac{M \delta M_4 \varepsilon ^4}{\rho^6} 
     +\mathcal{O}\left(\varepsilon^6\right).
\end{align}
\end{widetext}   


\section{The Johannsen metric  and the expansions of $\Omega$ and $M/r_0$}
\label{sec:Jexpansions}

Using Kerr-like coordinates $\{t, r,\theta,\varphi\}$, the separable Johannsen metric is given by~\cite{Johannsen:2013pca}
\begin{align}
g_{tt}  & = - \frac{\tilde{\Sigma}} {N} \left (\, \Delta -a^2 A_2^2  \sin^2\theta \, \right ), 
\\ 
g_{t\varphi}  &=  - \frac{a \tilde{\Sigma}} {N} \sin^2\theta \left [\, (r^2+a^2) A_1 A_2 -\Delta \, \right ],
\\ 
g_{\varphi\varphi}  &=  \frac{\tilde{\Sigma}} {N}  \sin^2\theta \left [\, (r^2+a^2)^2 A_1^2 -a^2 \Delta \sin^2\theta \, \right ],  
\\
g_{rr}  &= \frac{\tilde{\Sigma}}{\Delta A_5}, \qquad g_{\theta\theta} = \tilde{\Sigma},
\label{Jmetric}
\end{align}
where $a=J/M^2$ is the dimensionless spin parameter and
\begin{align}
\Delta &= r^2 -2 M r + a^2,
\\ 
N &= \left  [\, (r^2+a^2) A_1 -a^2 A_2 \sin^2\theta \, \right  ]^2, 
\\
\tilde{\Sigma} &= r^2 + a^2 \cos^2\theta + f(r). 
\end{align}
The deformation away from Kerr is encapsulated in the radial functions $\{A_1(r), A_2 (r), A_5 (r), f(r)\}$,
which can be written as power series of $1/r$. To leading order in the deviation from Kerr, these take the simple form
\begin{align}
A_1 &= 1 + \alpha_{13} \left (\frac{M}{r} \right )^3, \qquad  A_2  = 1 + \alpha_{22} \left (\frac{M}{r} \right )^2
\\
A_5  & = 1 + \alpha_{52} \left (\frac{M}{r} \right )^2, \qquad f  = \varepsilon_3 \frac{M^3}{r},
\end{align}
where $ \alpha_{13},  \alpha_{22},  \alpha_{52}, \varepsilon_3$ are the constant deformation  parameters. 

Considering a test body's circular orbit of radius $r_0$, the expansion of the angular frequency $\Omega$ in $M/r_0$ and the  deformation parameters
(with $\epsilon$ used as a collective bookkeeping parameter) is
\begin{widetext}
\begin{align}
\Omega &= \left ( \frac{M}{r_0^3} \right )^{1/2} \left [ 1 - a  \left ( \frac{M}{r_0} \right )^{3/2} +  a^2 \left (\frac{M}{r_0} \right )^3
- a^3 \left ( \frac{M}{r_0} \right )^{9/2} + \epsilon \left \{ \frac{3}{4} (2  \alpha_{13} - \varepsilon_3)  \left (\frac{M}{r_0} \right )^2  
-   a \alpha_{22}  \left ( \frac{M}{r_0} \right )^{5/2} 
\right. \right.
\nonumber \\
& \left. \left. +  \left ( \frac{9}{4} \varepsilon_3 -4  \alpha_{13} \right ) \left ( \frac{M}{r_0} \right )^3
 -\frac{3}{ 2}  a  \alpha_{13}  \left ( \frac{M}{r_0} \right )^{7/2} 
+ \frac{1}{2} (5  \alpha_{13} + 14  \alpha_{22})  a^2   \left ( \frac{M}{r_0} \right )^4  
\right. \right.
\nonumber \\
& \left. \left.  -  \frac{1}{2}  a  \left ( 4 a^2 \alpha_{22}  -16  \alpha_{13}  + 9 \varepsilon_3 \right )  \left ( \frac{M}{r_0} \right  )^{9/2} \right \} 
- \frac{9 \epsilon^2}{32} (\varepsilon_3 -2 \alpha_{13} )^2  \left ( \frac{M}{r_0} \right )^{4} + {\cal O} \left (\epsilon^3, \frac{M^5}{r_0^5} \right ) \right ].
\label{Omexpan}
\end{align}
\end{widetext}
The `inverse' expansion of $M/r_0$ in  $ v^3 = M \Omega$ and $\epsilon$ is
\begin{widetext}
\begin{align}
\frac{M}{r_0} &= v^2 + \frac{2}{3}  a v^5 + \frac{1}{2} \epsilon ( \varepsilon_3 -2 \alpha_{13})  v^6 + \frac{2}{3} \epsilon a \alpha_{22}  v^7 
+  \left [ \frac{5}{9} a^2 + \frac{\epsilon}{6M} (16\alpha_{13} - 9\varepsilon_3) \right ]  v^8 
+ \epsilon  a ( 2\varepsilon_3 -3 \alpha_{13} ) v^9 
\nonumber \\
& + \frac{1}{9} \left [ 9 \epsilon^2 ( \varepsilon_3 -2\alpha_{13} )^2 - \epsilon a^2 ( 15 \alpha_{13} + 16 \alpha_{22} )   \right ] v^{10}  
+ {\cal O} \left (\epsilon^3, v^{11} \right ).
\label{vexpan}
\end{align}
\end{widetext}


\bibliography{biblio.bib}

\end{document}